\documentclass[12pt]{article}
\usepackage[latin9]{inputenc}
\usepackage[margin=1in]{geometry}
\usepackage{amssymb}
\usepackage{amsmath, mathtools, amsthm}
\usepackage{algorithm}
\usepackage{algpseudocode}
\usepackage{subfiles}
\usepackage{xparse}
\usepackage{float}
\usepackage{bm}
\usepackage{graphicx}
\usepackage[unicode=true,
 bookmarks=false,
 breaklinks=false,pdfborder={0 0 1},colorlinks=false]
 {hyperref}
\usepackage{cleveref, autonum} 
\usepackage{prettyref}
\usepackage[sort&compress,elide]{natbib}
\usepackage{macros}
\usepackage{multirow}
\usepackage{color}
\usepackage[compact]{titlesec}

\newtheorem{theorem}{\protect\theoremname}[section]

\newtheorem{lemma}[theorem]{\protect\lemmaname}

\newtheorem{assumption}{\protect\assumptionname}
\makeatother

\providecommand{\lemmaname}{Lemma}
\providecommand{\theoremname}{Theorem}
\providecommand{\assumptionname}{Assumption}

\makeatletter
\let\oldabstract\abstract
\let\oldendabstract\endabstract
\makeatletter
\renewenvironment{abstract}
{
\oldabstract}
{\oldendabstract}
\makeatother

\newcommand{\blind}{1}

\titlespacing{\section}{0pt}{2ex}{1ex}

\begin{document}

\if1\blind
{
  \title{\bf Functional Group Bridge for Simultaneous Regression and Support Estimation}
    \author{{\small Zhengjia Wang$^{1}$, John F. Magnotti$^{2}$, Michael S. Beauchamp$^{2}$, and Meng Li$^{1}$}\\
    \vspace{0.1cm} \\
    {\small \it $^{1}$ Department of Statistics, Rice University} \\
    {\small \it $^{2}$ Department of Neurosurgery, } \\
    {\small \it University of Pennsylvania Perelman School of Medicine}
    }
    \date{}
  \maketitle
  \vspace{-0.3cm}
} \fi

\if0\blind
{
  \bigskip
  \bigskip
  \bigskip
  \begin{center}
    {\LARGE\bf Functional Group Bridge for Simultaneous Regression and Support Estimation}
\end{center}
  \medskip
} \fi

\bigskip

\label{firstpage}

\begin{abstract}
This article is motivated by studying differential brain activities to multiple experimental condition presentations in intracranial electroencephalography (iEEG) experiments. Contrasting effects of experimental conditions are often zero in most regions and non-zero in some local regions, yielding locally sparse functions. Such studies are essentially a function-on-scalar regression problem, with interest being focused not only on estimating nonparametric functions but also on recovering the function supports. We propose a weighted group bridge approach for simultaneous function estimation and support recovery in function-on-scalar mixed effect models, while accounting for heterogeneity present in functional data. We use B-splines to transform sparsity of functions to its sparse vector counterpart of increasing dimension, and propose a fast non-convex optimization algorithm using nested alternative direction method of multipliers (ADMM) for estimation. Large sample properties are established. In particular, we show that the estimated coefficient functions are rate optimal in the minimax sense under the $L_2$ norm and resemble a phase transition phenomenon. For support estimation, we derive a convergence rate under the $L_{\infty}$ norm that leads to a selection consistency property under $\delta$-sparsity, and obtain a result under strict sparsity using a simple sufficient regularity condition. An adjusted extended Bayesian information criterion is proposed for parameter tuning. The developed method is illustrated through simulations and an application to a novel iEEG dataset to study multisensory integration. 
\end{abstract}

\section{Introduction}
\label{s:intro}

Functional data analysis (FDA) is routinely encountered in modern applications due to the rapid advancement of new techniques to collect high-resolution data that can be viewed as curves; see~\cite{ramsay2005fda, morris2015functional, wang2016functional} for a comprehensive treatment. An overwhelming focus has been on nonparametric estimation of the underlying functions. However, shape constraints arise naturally in modern applications. 
{{One such example is \textit{local sparsity}, i.e., the function is exactly zero on subregions, in contrast to global sparsity that refers to a zero function.}} Local sparsity is a crucial characteristic for a nonparametric method to be interpretable in a variety of applications, and the estimation of the support as well as the function itself is of primary interest. This article aims to develop a flexible method with efficient implementation and theoretical guarantees for simultaneously estimating and recovering the support of locally sparse functions. 

\begin{figure}
	\centering
	\includegraphics[width=5in]{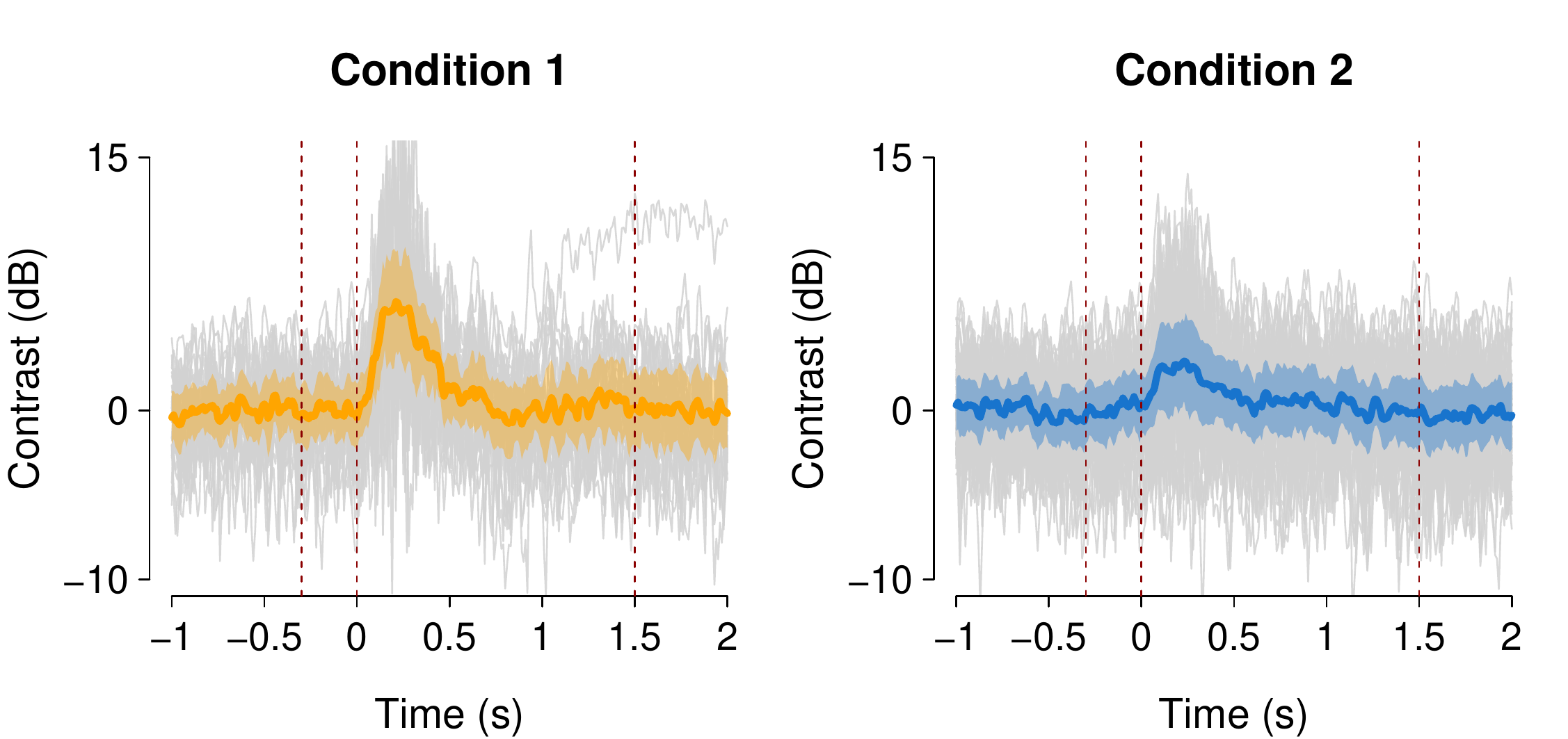}
	\caption{Baseline corrected response of iEEG high-gamma power under auditory-only (left) and audiovisual (right) conditions. The data are obtained by decomposing the measured voltage signal into frequency space, converting to Decibel unit, calibrated to baseline (from $-1$ to $-0.3$ seconds), and then taking the average power in the 70-150Hz range. Each individual grey line is one trial. The bold solid line is the mean responses, and the shaded area around the mean is a pointwise 95\% confidence interval. \label{fig:intro-demo}}
\end{figure}

Our motivation stems from neuroscientific studies using human intracranial electroencephalography (iEEG) data. {{iEEG is an emerging invasive method that offers anatomically precise measurements of human brain activity with electrodes placed on or implanted in the human brain, leading to excellent temporal resolution data and high signal-to-noise ratios \citep{lachaux2012high,kaiju2017high}}}. In most iEEG experiments, {{participants are presented with multiple experimental conditions. The brain response to each condition is recorded, and the experimenter wishes to know whether and how they differ}}. The contrast of brain activities is expected to be locally sparse (zero at certain period of time), and detecting non-sparse regions is of substantial interest to neuroscientists in addition to estimating the coefficient functions. For example, Figure \ref{fig:intro-demo} shows iEEG data from 
an audiovisual speech perception task \citep{ozker2018frontal} under two experimental conditions, namely, ``auditory-only'' and ``audiovisual''. In this study, the goal is to understand how the brain responds to auditory and visual stimuli through analyzing differential brain activities to these two conditions. Large trial-to-trial variation necessitates the use of statistical inference to automate the extraction of both population trajectories and supports of underlying brain activities. 

Over the past several decades, there has been an extensive literature on sparsity. This leads to a rich menu of methods in the context of regularization-based variable selection with the parameter space being \textit{sparse vectors}, including the Lasso~\citep{tibshirani1996regression}, 
minimax concave penalty (MCP)~\citep{zhang2010nearly}, 
and bridge regression~\citep{frank1993statistical}, to name just a few; see \cite{fan2010selective} for a review. Such concepts have been extended to grouped variable selection and nonparametric \textit{sparse functions}.
{{Indeed, coupled with basis expansions, sparse coefficient vectors regularize the estimated function and lead to globally sparse coefficient functions via grouped sparsity. Along this line, \cite{barber2017} extended group Lasso to functional data, and \cite{chen2016variable} adopted group MCP, both achieving variable selection and parameter estimation.}} 

However, comparatively little work has been done for functions with local sparsity. {{There is a related literature on utilizing various penalties to improve interpretability, for example,  \cite{james2009functional,zhou2013functional,wang2015functional,lin2017locally}}}.
These methods, however, rely on scalar responses and are not suitable for the motivating iEEG studies, where the response is functional and covariates are scalar; hereafter we refer to such cases as \textit{function-on-scalar regression}. Function-on-scalar regression poses unique challenges to methodological and theoretical developments. In particular, the intrafunctional dependence of responses is vital, and the large sample properties may be more intricately determined by the sample size (number of subjects) as well as the sampling frequency of individual trajectories of each subject. {{In what follows we focus our attention on longitudinal data over time that out motivating example corresponds to.}}

In this paper, we propose a method for simultaneous regression and support estimation for function-on-scalar regression, where the underlying functions are locally sparse. We adopt group bridge estimators coupled with B-splines to recover the sparse pattern of functional regression coefficients. {{Unlike the group Lasso and Lasso, group bridge penalty provably achieves variable selection in linear models at the group and individual level simultaneously \citep{huang2009group}. This is particularly well suited for locally sparse functions by inducing exactly zero regions through grouping basis coefficients that contribute to the function at each time point, while maintaining parsimony and regularization of basis coefficients via selection at the individual level. Following our use of B-splines and grouping structure of variables, other penalties can also be considered if they ensure both group-level and individual-level sparsity; however, these methods have not been extended to function-on-scalar regression.}} 

The proposed method does not require Gaussian assumptions and {{allows flexible heterogeneous correlation structures through random effects that possibly depend on various experimental phases.}} This leads to a novel \textit{weighted functional group bridge} approach for function-on-scalar mixed effect models. 
On the algorithmic front, we introduce a nested alternating direction method of multipliers (ADMM) algorithm with ``warm-start'' and ``early stopping'' to speed up the computation. On the theoretical front, we establish a range of large sample properties, including rate optimality in the minimax sense for regression and selection consistency for support estimation. 
Although under different models, our theory relies on substantially simplified assumptions than the existing literature to regularize regression functions, notably Assumption~\ref{assumption:integral}, facilitating interpretability. We allow flexible sampling designs including the case when 
the number of time points grows faster than the sample size, which is better suited for iEEG studies. In an application to the aforementioned iEEG experiment, our results complement previous studies by showing that multisensory interactions are a powerful modulator of activity throughout the speech perception network~\citep{ozker2018frontal, karas2019cross}.

The rest of the article is organized as follows. Section \ref{s:methodology} introduces our model, the proposed estimator, and the optimization algorithm. Section \ref{s:asymptotic-properties} provides asymptotic properties. Section \ref{s:simulation} contains simulations, followed by an application to iEEG analysis in Section \ref{s:application}. The proposed method is implemented in the R package {\verb!spfda!}, available on CRAN. The Appendix include all proofs and additional simulations. 

\section{Methodology}
\label{s:methodology}

\subsection{Model}

Suppose a sample of $n$ functional signals $\bracketlg[s]{{y}_{i}(\time)}_{i = 1}^n$ is observed on a compact time set $\mathcal{T}$. We assume without loss of generality $\mathcal{T}=\bracketmd{0,1}$. The linear function-on-scalar mixed effect model assumes 
\begin{equation} \label{eq:FDA.model} 
{y}_{i}(\time)=\sum_{j=1}^{p}x_{ij}{\beta}_{j}(\time)+ {\theta}_{i}(\time) + {\epsilon}_{i}(\time), 
\end{equation}
where $x_{ij}$ is 
a scalar predictor, ${\beta}_j(\time)$ is a fixed effect function, ${\theta}_{i}(\time)$ is a zero mean random effect with {{covariance kernel $\Sigma_{\bm{\theta}}(\time,\time')$ that captures the within-curve dependence}}, and the ${\epsilon}_{i}(\time)$ is the measurement error process independent of ${\theta}_{i}(\time)$ with zero means and covariance kernel $\Sigma_{\bm{\epsilon}}(\time,\time')=\sigma^{2}\delta_{t, t'}$; here $\delta_{t, t'}$ is the Kronecker delta. 
We assume ${\beta}_j(\time)$ is smooth and locally sparse on the time domain. We propose to use phase-dependent random effects to account for heterogeneous dependence structures in various stages of an experiment, such as resting phase, trial onset, and stimuli offset. In particular, we partition $\mathcal{T}$ into a union of disjoint intervals $\bracketlg[s]{\mathcal{T}^{\text{pa}}}_{\text{pa}=1}^{P}$, where each $\mathcal{T}^{\text{pa}}$ corresponds to a stage of the experiments, and random effects ${\theta}_{i}(\time)$ are smooth within each phase. Note this generalizes traditional mixed effect models where $P$ is typically set to one.

We use B-splines $\bracketlg{{\phi}_{k}(\time)}_{k=1}^{K}$ to approximate each fixed effect function, that is, ${\beta}_j(\time) = \sum_{k = 1}^K {\gamma}_{jk} {\phi}_k(\time) + {R}_{j}(\time)$, where ${R}_j(\time)$ is the approximation error. In addition to sharp approximation bounds to smooth functions, B-splines are particularly well suited for sparse functions as they are locally supported and thus transfer sparsity in ${\beta}(\time)$ to a sparse $p$ by $K$ matrix $\bm{\gamma} = \{{\gamma}_{jk}\}_{p\times K}$.
Let a non-negative integer $d$ be the degree of B-splines and define the knots of length $K+d+1$ be $\time_{\tidx_{1}}=\dots=\time_{\tidx_{d}}=0=\time_{\tidx_{d+1}}<\time_{\tidx_{d+2}}<\dots<\time_{\tidx_{K}}=1=\time_{\tidx_{K+1}}=\dots=\time_{\tidx_{K+d+1}}$. Then B-splines are defined recursively \citep{de1978practical} as follows:
\begin{gather}
{\phi}_{k,1}(\time) = \mathbf{1}_{[\time_{ \tidx_{k} }, \time_{ \tidx_{k+1} })}(t)
\label{eq:spl-def1}
,\quad
{\phi}_{k,d+1}(\time) = \frac{\time-\time_{\tidx_{k}}}{ \time_{\tidx_{k+d}} - \time_{\tidx_{k}} }{\phi}_{k,d}(\time) + \frac{ \time_{\tidx_{k+1}} - t }{ \time_{\tidx_{k+q-1}} - \time_{\tidx_{k+1}} }{\phi}_{k+1,d}(\time),  \label{eq:spl-def2}
\end{gather}
where $k = 1, \ldots, K$. 
B-splines of order $q = d + 1$ are ${\phi}_{k}(\time) = {\phi}_{k,q}(\time)$, and we typically choose $q\geq 4$.

In practice, functional data are observed at discrete time points. Let  $\mathcal{T}_{0}=\bracketlg{\time_{\tidx}}_{\tidx=1}^{T} \subseteq \mathcal{T}$ be the set of time points at which ${y}_{i}(\time)$ is observed. Each partition set $\mathcal{T}_{0}^{\text{pa}}$ is defined as $\mathcal{T}_{0}\cap \mathcal{T}^{\text{pa}}$. Let
$\bm{Y}=\bracketlg{y_{i \tidx}}_{n\times T}, \bm{\theta}=\bracketlg{{\theta}_{i\tidx}}_{n\times T}, \bm{E}=\bracketlg[s]{e_{it}}_{n\times T}$ be the discretized responses, random effects, and random noise observed at $\mathcal{T}_{0}$, respectively, and {{$\bm{X}=\bracketlg{x_{ij}}_{n\times p}$ the design matrix}}, $\bm{B}=\bracketlg{{B}_{k}^{\tidx}}_{K\times T}$ the basis functions $\bracketlg{{\phi}_{k}(\time)}_{k=1}^{K}$ evaluated on $\mathcal{T}_{0}$, and $\bm{R}=\bracketlg{{R}_{j}(\time_{\tidx})}_{p\times T}$ the corresponding approximation error.
Then Model~\eqref{eq:FDA.model} can be written as
\begin{equation}
\bm{Y}=\bm{X}\bm{\gamma} \bm{B} + \bm{X}\bm{R} + \bm{\theta} + \bm{E},  \label{eq:2-4}
\end{equation}
where $\bm{\theta}$ and $\bm{E}$ have covariance $\bm{\Sigma}_{\bm{\theta}}$ and $\bm{\Sigma}_{\bm{\epsilon}}=\sigma^{2}\bm{I}$ that are discretized $\Sigma_{\bm{\theta}}(\time,\time')$ and $\Sigma_{\bm{\epsilon}}(\time,\time')$ on $\mathcal{T}_{0}$.

In what follows, we use subscript to index rows; for example, $\bm{y}_{i}, \bm{X}_i, \bm{\gamma}_j, \bm{B}_k$ are the corresponding rows of $\bm{Y}, \bm{X}, \bm{\gamma}, \bm{B}$, respectively. We use  $\bm{B}^{(\tidx)}$ to denote the $\tidx^{th}$ column of $\bm{B}$, and $\bm{X}^{(j)}$ to denote the $j^{th}$ column of $\bm{X}$. All vectors are column vectors.

\subsection{Estimation: Weighted functional group bridge}\label{sec:group.penalty}

We propose a weighted functional group bridge approach to estimate $\bm{\gamma}$: 
\begin{equation}\label{eq:gamma.hat} 
\widehat{\bm{\gamma}} =\argmin{\bm{\gamma}} \; L(\bm{\gamma}; \lambda, \alpha, \bm{W}) = \argmin{\bm{\gamma}} \; \{f(\bm{\gamma}; \bm{W})+\lambda g(\bm{\gamma}; \alpha)\},
\end{equation}
where $f(\bm{\gamma}; \bm{W})=\frac{1}{2}\sum_{i=1}^{n}\|(\bm{y}_{i}^{T}-\bm{X}_{i}^{T}\bm{\gamma} \bm{B})\bm{W}\|_{2}^{2}$ is the squared error loss with each observation weighted by a $T\times T$ matrix $\bm{W}$, and $\lambda g(\bm{\gamma}; \alpha)$ is a penalty term to encourage sparsity on $\bm{\gamma}$ and ${\beta}(t)$ with tuning parameters $\lambda \geq 0$ and $\alpha > 0$. Each coefficient function ${\beta}_j(\time)$ for $j = 1, \ldots, p$ is estimated by $\widehat{{\beta}}_j(\time)=\sum_{k = 1}^K \widehat{{\gamma}}_{jk} {\phi}_{k}(\time)$. 

{{We use a \textit{group bridge} penalty for $g(\bm{\gamma}; \alpha)$ to achieve sparsity in both $\bm{\gamma}$ and ${\beta}_j(t)$: 
\begin{equation}
g(\bm{\gamma}; \alpha) =\sum_{j=1}^{p}\sum_{\tidx=1}^{T}g_{j,\tidx}(\bm{\gamma}; \alpha), \quad 
g_{j,\tidx}(\bm{\gamma}; \alpha) = \left\{\sum_{k: \bm{B}_k^{(m)} \neq 0} \abs{{\gamma}_{jk}}\right\}^{\alpha},
\end{equation}
which decomposes $g(\bm{\gamma}; \alpha)$ into $pT$ groups. 
Within each group, the $L_1$ penalty on a subset of $\bm{\gamma}$ leads to sparse estimates $\widehat{\bm{\gamma}}_j$~\citep{tibshirani1996regression}. 
At the group level, if $g_{j,\tidx}(\bm{\gamma}; \alpha) = 0$, then $\abs{{\beta}_{j}(\time_{\tidx})}=\abs[s]{\bm{\gamma}_{j}^{T}\bm{B}^{(m)}}=0$. Hence, group-level sparsity on $g_{j,\tidx}(\bm{\gamma}; \alpha)$ leads to sparse ${\beta}_j(\time_{\tidx})$ at $\time_{\tidx}$.
\cite{knight2000asymptotics} shows bridge estimators with $\alpha\in (0,1]$ combine variable selection and parameter estimation for sufficiently large $\lambda$. 
To achieve sparsity in both $\bm{\gamma}$ and ${\beta}_j(t)$, we propose to use $\alpha \in (0, 1)$ as it has the appealing property to select variables at the individual and group level simultaneously~\citep{huang2009group}. Note that $g(\bm{\gamma}; \alpha)$ becomes the $L_1$ penalty when $\alpha = 1$, which does not explicitly point to group-level sparsity that is critical to ensure that ${\beta}_j(t)$ is exactly zero at some $t$.}} Nevertheless, the developed algorithm in the following section is applicable for both $\alpha \in (0, 1)$ and $\alpha = 1$, and we further compare these two variants in simulations. To use a compact notation for $g_{jm}(\bm{\gamma}; \alpha)$, we indicate with $\mathbf{1}\{\bm{B}^{(\tidx)}\} = \big[\mathbf{1}\{{B}_{k}^{(\tidx)}\}\big]$ the indicator function evaluated on $\bm{B}^{(\tidx)}$, a column vector whose $k$th element $\mathbf{1}\{{B}_{k}^{(\tidx)}\}=1$ if ${B}_{k}^{(\tidx)}\neq 0$, and zero otherwise. Then 
\begin{equation}
g_{j,\tidx}(\bm{\gamma}; \alpha) =
\bracketmd[\bigg]{\sum_{k=1}^{K}\abs{{\gamma}_{jk}}\cdot  \mathbf{1}\bracketlg[s]{{B}_{k}^{(\tidx)}}}^{\alpha} = \|\mathbf{1}\bracketlg[s]{\bm{B}^{(\tidx)}}\odot\bm{\gamma}_{j}\|_{1}^{\alpha}, 
\end{equation}
where $\norm{\cdot}_{1}$ is the $L_{1}$ norm, and $\mathbf{1}\bracketlg[s]{\bm{B}^{(\tidx)}}\odot\bm{\gamma}_{j}$ is the element-wise multiplication between $\mathbf{1}\bracketlg[s]{\bm{B}^{(\tidx)}}$ and $\bm{\gamma}_{j}$.

{{One needs to specify $(\lambda, \alpha, \bm{W})$ in the objective function $L(\bm{\gamma}; \lambda, \alpha, \bm{W})$. 
In the next sections, to ease exposition we assume these parameters are given.}} 

As such, 
in the sequel we omit $(\alpha, \lambda, \bm{W})$ in $L(\bm{\gamma};\lambda,\alpha,\bm{W}), f(\bm{\gamma};\bm{W})$ and $g(\bm{\gamma};\alpha)$ and instead use $L(\bm{\gamma}), f(\bm{\gamma})$, and $g(\bm{\gamma})$, respectively, when it does not cause confusion. {We will introduce fully data-driven methods to select $\bm{W}$ and $(\alpha, \lambda)$ in Section~\ref{s:parameter-tuning}.}

\subsection{Optimization: Nested ADMM algorithm}
We first recast the minimization of $L(\bm{\gamma})$ into an iterative Lasso problem as in \cite{huang2009group}. In particular, we embed $g(\bm{\gamma})$ in a carefully chosen higher dimensional space then link the solution back to $g(\bm{\gamma})$ through a particular path.
Denote the expanded surface as
\begin{gather}
    S\bracketsm{\bm{\gamma},\bm{\zeta}}  =\sum_{j,\tidx}s_{j,\tidx}\bracketsm{\bm{\gamma},\bm{\zeta}}, \quad \bm{\zeta}=\{{\zeta}_{j,\tidx}\}, \\
s_{j,\tidx}(\bm{\gamma},\bm{\zeta})  =\frac{\alpha^{\alpha}}{(1-\alpha)^{\alpha-1}}{\zeta}_{j,\tidx}^{1-\frac{1}{\alpha}}\|\mathbf{1}\bracketlg[s]{\bm{B}^{(\tidx)}}\odot\bm{\gamma}_{j}\|_{1}
+\alpha^{\alpha}(1-\alpha)^{1-\alpha}{\zeta}_{j,\tidx}.
\end{gather}
The original non-convex problem in Equation~\eqref{eq:gamma.hat} can be solved by finding the minimizer for
\begin{gather}
\argmin{\bm{\gamma},\bm{\zeta}}\  L_{S}(\bm{\gamma},\bm{\zeta}) \quad 
\text{subject to} \quad  \bm{\zeta}=\bm{\zeta}(\bm{\gamma}) \text{ and } \bm{\zeta} > 0, \label{eq:modified-objective}\\
\text{where } 
L_{S}(\bm{\gamma},\bm{\zeta})=f(\bm{\gamma})+S(\bm{\gamma},\bm{\zeta})\label{eq:modobj}
, \text{ and } 
{\zeta}_{j,\tidx}(\bm{\gamma})  =\bracketsm[\Big]{\frac{1-\alpha}{\alpha}}^{\alpha}\|\mathbf{1}\bracketlg[s]{\bm{B}^{(\tidx)}}\odot\bm{\gamma}_{j}\|_{1}^{\alpha}.
\end{gather}
We carry out the optimization by iteratively updating $\bm{\zeta}$ with fixed $\bm{\gamma}=\bm{\gamma}^{\mathrm{old}}$ through the definition $\bm{\zeta} = \bm{\zeta}(\bm{\gamma})$,  
and updating $\bm{\gamma}$ by solving a Lasso problem with fixed $\bm{\zeta}=\bm{\zeta}^{\mathrm{new}}$.

We propose to solve the iterative Lasso problem using nested alternating direction method of multipliers, or ADMM~\citep{boyd2011distributed}.  Algorithm \ref{alg:1} details the nested ADMM algorithm, in which we use common notations for matrices. For any matrix $\bm{A}$, $\bm{A}^{+}$ is the positive part, and $\bm{A}^{-}$ is the negative part. The operator $\diag(\cdot)$ extracts the diagonal elements of a square matrix into a vector, and expands a vector to a diagonal square. The operator $\oslash$ between two matrices defines element-wise division. 

{We use the ridge regression estimate $\widehat{\bm{\gamma}}_{\rm ridge}$ that minimizes \eqref{eq:gamma.hat} with $g(\bm{\gamma}; \alpha) = \sum_j \| \bm{\gamma}_j \|_2^2$ and tuning parameter $\lambda_{\rm ridge}$ as a closed-form ``warm-start".}  Compared to LARS~\citep{efron2004least} adopted by \cite{huang2009group}, warm started ADMM can significantly improve the performance: If initialized near the solution, ADMM converges faster to modest accuracy within a few steps~\citep{boyd2011distributed, majzoobi2018analysis}. To fully take advantage of this property, we use few ADMM steps within each iteration without checking or waiting till full convergence, leading to ``early stopping''. The solutions at each iteration are used as warm-starts for the next iteration, further speeding up the convergence. {{Although the partial derivative $\partial g(\bm{\gamma})/\partial {\gamma}_{jk}$ near zero diverges, zero does not tend to be an absorbing state with an increased augmented Lagrangian parameter $\rho$ in Algorithm \ref{alg:1}. In addition, 
the adopted dense initialization via ridge regression and early stopping in ADMM help prevent the coefficients from entering zeros at early stage,}} leaving enough iterations for the coefficients to prioritize fitting before becoming sparse. {{In numerical experiments not reported here, we have found that the proposed algorithm outputs similar estimates with radically different initial values, indicating robustness. Nevertheless, we recommend to use the ridge regression estimate for faster convergence.}} 

{{We use the following default settings in our numerical experiments, unless otherwise stated. We use 5-fold cross validation to select $\lambda_{\rm ridge}$ in ridge regression. The number of iterations $S_{1}$ is 20, and the number of ADMM steps per iteration is $S_{2} = 50$. }}We increase $\rho$ exponentially by letting $\rho = e^{4s / S_{2} - 1}$ where $s$ is current ADMM step.  
\begin{algorithm}[ht]
\caption{Nested ADMM solver for the case $\alpha\in(0,1)$}\label{alg:1}
\begin{algorithmic}[1]
\Procedure{}{$\bm{X},\bm{Y},\bm{B},\lambda, \alpha, \bm{W}, S_{1}, S_{2}$}\Comment{Inputs}
\State Initialize $\bm{\gamma}\leftarrow \bm{\gamma}^{(0)}\leftarrow \hat{\bm{\gamma}}_{\rm ridge}$, $\bm{E}\leftarrow \bm{Y}-\bm{X}\bm{\gamma} \bm{B}$;
\State SVD decomposition: $\bm{X} =\bm{U}_{\bm{X}}\bm{D}_{\bm{X}}\bm{V}_{\bm{X}}^{T}$, $\bm{B}\bm{W} =\bm{U}_{\bm{B}}\bm{D}_{\bm{B}}\bm{V}_{\bm{B}}^{T}$;
\State Assign $\bm{V} = \bm{V}_{\bm{X}}^{T}$, $\bm{U} = \bm{U}_{\bm{B}}$, $\bm{Z} = \diag(\bm{D}_{\bm{X}}^{2})\diag(\bm{D}_{\bm{B}}^{2})^{T}+\rho\mathbf{1}_{p}\mathbf{1}_{K}^{T}$ ;
\For{$v$ in $1,2,\dots,S_{1}$} \Comment{Macro-loop: iterative Lasso}
    \State Calculate $\bm{D}^{(v)} = \{{D}_{jk}^{(v)}\},$ where
        ${D}_{jk}^{(v)} = \alpha^{\alpha}(1-\alpha)^{1-\alpha}{\zeta}_{j,\tidx}^{1-\frac{1}{\alpha}}(\bm{\gamma}^{(v-1)})\cdot\mathbf{1}\bracketlg[s]{{B}_{k}^{(\tidx)}}$; 
    \State Initialize $\bm{\eta}^{(0)}=\bm{\gamma}^{(v-1)}$, $\bm{\xi}^{(0)}=\bm{\gamma}^{(v-1)},$ $\bm{\psi}^{(0)} = \bm{\xi}^{(0)}-\bm{\eta}^{(0)} = \bm{0}$;
    \For{$s$ in $1,2,\dots,S_{2}$} \Comment{ADMM loop, fixed steps}
        \State $\bm{\phi}\leftarrow \bm{V}\bm{X}^{T}\bm{Y}\bm{W}(\bm{B}\bm{W})^{T}\bm{U}-\bm{\psi}^{(s-1)}+\rho \bm{V}\bm{\eta}^{(s-1)}\bm{U}$
        \State  $\bm{\xi}^{(s)}\leftarrow \bm{V}^{T}\left(\bm{\phi}\oslash \bm{Z}\right)\bm{U}^{T}$; 
        \State  $\bm{\eta}^{(s)}\leftarrow\bracketlg[\big]{\bm{\xi}^{(s)}+\frac{1}{\rho}\bm{V}^{T}\bm{\psi}^{(s-1)}\bm{U}^{T}-\frac{\lambda}{\rho}\bm{D}^{(v)}}^{+}-\bracketlg[\big]{\bm{\xi}^{(s)}+\frac{1}{\rho}\bm{V}^{T}\bm{\psi}^{(s-1)}\bm{U}^{T}+\frac{\lambda}{\rho}\bm{D}^{(v)}}^{-}$;
        \State  $\bm{\psi}^{(s)}  \leftarrow\bm{\psi}^{(s-1)}+\rho \bm{V}\bracketlg[s]{\bm{\xi}^{(s)}-\bm{\eta}^{(s)}}\bm{U}$;
    \EndFor
    \State Assign $\bm{\gamma}\leftarrow \bm{\gamma}^{(v)}\leftarrow \bm{\eta}^{(S_{2})}$;
    \State Recall $\widehat{\bm{Y}}=\bm{X}\bm{\gamma}^{(v)}\bm{B}$, calculate $\widehat{\bm{E}}=\bm{Y}-\widehat{\bm{Y}}$;
    \If{ $\|\widehat{\bm{E}} \bm{W}\|_{2}$ is stable }{ Break the loop} \Comment{Exit}
    \EndIf
\EndFor
\State \textbf{Return} $\bm{\gamma}$;
\EndProcedure
\end{algorithmic}
\end{algorithm}

Not that by setting $S_{2}=1$ and forcing ${D}^{(v)}_{jk}=1$ at each iteration, Algorithm~\ref{alg:1} also solves the case when $\alpha=1$. 

\subsection{Variance Estimation} \label{s:variance}

Algorithm \ref{alg:1} solves an augmented Lagrangian problem $L^\text{aug}$:
\begin{equation}
    L^\text{aug}(\bm{\xi}, \bm{\zeta}, \bm{\psi}, \bm{\eta}; \bm{W}, \lambda, \alpha, \rho) = L_S(\bm{\xi}, \bm{\zeta}) 
    + \bracketlg{ \vec{(\bm{\psi})}^{T} \vec{\bracketsm{\bm{V}\bm{\xi} \bm{U} - \bm{V}\bm{\eta} \bm{U}}} } + \frac{\rho}{2}\norm{\bm{V}(\bm{\xi}-\bm{\eta})\bm{U}}_{2}^{2},
\end{equation}
where $\vec(\bm{A})$ vectorizes $\bm{A}$ by stacking columns of $\bm{A}$. 
The dual feasibility in the ADMM optimality condition yields
\begin{gather}
    \{\bm{V}\bm{\xi}^{(s)} \bm{U}\} \odot \bm{Z} = \bm{V}\bm{X}^{T}\bm{Y}\bm{W}\bm{W}^{T}\bm{B}^{T}\bm{U} - \bm{\psi}^{(s-1)} +\rho \bm{V}\bm{\eta}^{(s-1)}\bm{U} \label{eq:dual-1}\\
    \mathbf{0} = \bracketmd[\Big]{-\frac{\lambda}{\rho} D(\bm{\zeta})\text{sign}\bracketlg[\big]{\bm{\xi}^{(s)}} + \frac{1}{\rho}\bm{V}^{T}\bm{\psi}^{(s-1)}\bm{U}^{T} - \bm{\eta}^{(s)}} \odot \text{sign}\bracketlg[\big]{\bm{\xi}^{(s)}}. \label{eq:dual-2}
\end{gather}
When Algorithm \ref{alg:1} converges, $\rho (\bm{\eta}^{(s)}-\bm{\eta}^{(s-1)})\rightarrow \mathbf{0}$ and $\bm{\xi}^{(s)} \rightarrow \widehat{\bm{\gamma}}$. Hence, for $(j,k)$ such that $\widehat{{\gamma}}_{jk}\neq 0$, the preceding optimality conditions in \eqref{eq:dual-1} and \eqref{eq:dual-2} yield
\begin{equation}
    {\bm{X}^{(j)}}^T \bm{X}\widehat{\bm{\gamma}} \bm{B}\bm{W}\bm{W}^T\bm{B}_{k} + \lambda {D}_{jk}(\widehat{\bm{\zeta}})\widehat{{\gamma}}_{jk} 
    = {\bm{X}^{(j)}}^T \bm{X}\bm{\gamma} \bm{B}\bm{W}\bm{W}^T\bm{B}_{k} + {\bm{X}^{(j)}}^T \bm{E}\bm{W}\bm{W}^T\bm{B}_{k}. \label{eq:optimality}
\end{equation}
Denote $\bm{P}(j,k)=\vec\{\bm{X}^T\bm{X}^{(j)}\bm{B}_{k}^T\bm{W}\bm{W}^T\bm{B}^T+\lambda {D}_{jk} \bm{J}(j,k)\}$, where $\bm{J}(j,k)$ is sparse matrix with only its $(j,k)$ element being $1$, and let  $\bm{Q}(j,k)=\bm{X}^{(j)}\bm{B}_{k}^T\bm{W}\bm{W}^T$. Then $(\ref{eq:optimality})$ can be written as
\begin{equation}
    \vec\bracketlg{\bm{P}(j,k)}^T\vec(\widehat{\bm{\gamma}}) = \vec\bracketlg{\bm{Q}(j,k)}^T \vec(\bm{X}\bm{\gamma} \bm{B}+\bm{Z}\bm{\theta}+\bm{E}), \label{eq:gamma-implicit}
\end{equation}
for $(j,k)$ such that $\widehat{{\gamma}}_{jk}\neq 0$. Letting $\vec_{\lambda}(\widehat{\bm{\gamma}})$ be the sub-vector of $\vec(\widehat{\bm{\gamma}})$ on its support, that is, all elements in $\vec_{\lambda}(\widehat{\bm{\gamma}})$ are non-zeros, and likewise $\bm{P}_{\lambda}(j,k)$ be $\bm{P}(j,k)$ on its support, then we have $\vec\bracketlg{\bm{P}(j,k)}^T\vec(\widehat{\bm{\gamma}}) = \vec_{\lambda}\bracketlg{\bm{P}(j,k)}^T\vec_{\lambda}(\widehat{\bm{\gamma}})$. Bind $\vec_{\lambda}\bracketlg{\bm{P}(j,k)}$ by column for all $(j,k)$ such that $\widehat{{\gamma}}_{jk}\neq 0$, and denote it as $\bm{P}_{\lambda}$. It is easy to show that $\bm{P}_{\lambda}$ is a square matrix and further
$    \bm{P}_{\lambda}\vec_{\lambda}(\widehat{\bm{\gamma}}) = \bm{Q} \vec(\bm{X}\bm{\gamma} \bm{B}+\bm{Z}\bm{\theta}+\bm{E}),$ 
where $\bm{Q}$ is also a square matrix column-binded by $\vec\bracketlg{\bm{Q}(j,k)}$. 
Consequently, the covariance of $\widehat{\bm{\gamma}}$ is given by
\begin{equation}
    \text{Cov}\bracketlg{\vec_{\lambda}(\widehat{\bm{\gamma}})}=\bm{P}_{\lambda}^{-1}\bm{Q}\text{Cov}\bracketlg{\vec(\bm{Z}\bm{\theta}+\bm{E})}\bm{Q}^T\bracketsm{\bm{P}_{\lambda}^T}^{-1}. \label{eq:covariance}
\end{equation}

\subsection{Choice of weights and parameter tuning} \label{s:parameter-tuning}
We propose to use $\bm{W} \propto\bm{\Sigma}^{-1/2}$ to accounts for heterogeneity in the functional response, where $\bm{\Sigma}=\bm{\Sigma}_{\bm{\theta}} + \sigma^{2}\bm{I}$ is the covariance matrix of ${y}_{i}^{T}-\bm{X}_{i}^{T}{\beta}$.
When $\bm{\Sigma}_{\bm{\theta}}$ and $\sigma^{2}$ are unknown, we substitute using their estimates $\widehat{\bm{\Sigma}}_{\bm{\theta}}$ and $\widehat{\sigma}^{2}$ to derive $\widehat{\bm{\Sigma}}$ and subsequently $\bm{W}$. We propose to employ local linear regression~\citep{fan1993local, zhu2014spatially} to estimate $\bm{\Sigma}$, assuming the existence of the second-order derivative of ${\theta}_{i}(\time)$ within each phase $\mathcal{T}^{\text{pa}}$. Denote $\Delta_{\tidx,b}(\time)=(\time_{\tidx} - \time) / b$ and  ${\theta}'_{i,b}(\time)=b \cdot d{\theta}_{i}/d\time(\time)$, and choose a kernel function $K_{b}(\time)$ with the bandwidth parameter $b$ selected by minimizing the generalized cross-validation score~\citep{zhu2014spatially}. 
For each $\time\in\mathcal{T}^{\text{pa}}$, we estimate ${\theta}_{i}(\time)$ using a weighted least squares procedure \citep{fan1993local}:
\begin{align}
    \bracketlg[\Big]{\widehat{{\theta}}_{i}(\time),\widehat{{\theta}'}_{i}(\time) }^T
   =  \argmin{
        {\theta}_{i}(\time), {\theta}'_{i}(\time)
    }
    \sum_{
    \tidx: \time_{\tidx}\in \mathcal{T}_{0}^{\text{pa}}
    }
    &
    \Big[  
        {y}_{i}(t_{\tidx}) 
        - \sum_{j=1}^{p} x_{ij} \widehat{{\beta}}_{j}^{ols}(\time_{\tidx}) \\ 
        &\quad - \bracketlg[\big]{{\theta}_{i}(\time) + {\theta}'_{i}(\time)\Delta_{\tidx,b}(\time)} \Big]^{2}
    K_{b}(\time_{\tidx} - \time),
\end{align}
where $\widehat{{\beta}}_{j}^{ols}(\time_{\tidx})$ is ordinary lease square estimator of the $j^{th}$ coefficient at time $\time_{\tidx}$. We then calculate $\widehat{\bm{\Sigma}}_{\bm{\theta}}$ from the sample covariance of $\bracketlg[s]{\widehat{{\theta}}_{i}(\time_{\tidx})}_{i,\tidx}$, and $\widehat{\sigma}^2$ from the residuals ${y}_{i}(t_{\tidx}) - \sum_{j=1}^{p} x_{ij} \widehat{{\beta}}_{j}^{ols}(\time_{\tidx})- \widehat{{\theta}}_{i}(\time_{\tidx})$. {{As a passing comment, one may alternatively use $\bm{W} = \bm{I}$; this may lead to efficiency loss unless the functional dependence is homogeneous or approximately so, a special case of Model~\eqref{eq:FDA.model} without random effects. The developed methods and theory are applicable for this simplified model with such a choice of $\bm{W}$.}} 

We next move on to the tuning of $\alpha$ and $\lambda$ via an adjusted extended Bayesian information criterion (EBIC). We choose an equally spaced sequence of knots for B-splines. For given $K$,  
the EBIC proposed by \cite{chen2008extended} is
\begin{equation}
    EBIC_{\nu} = \frac{\norm{(\bm{Y}-\bm{X}\widehat{\bm{\beta}})\bm{W}}_2^2}{n\sigma^{2}} + T\log{(\sigma^{2})} + df \frac{\log{n}}{n} + \nu df \frac{\log{pK}}{n}, \label{eq:ebic}
\end{equation}
where $\widehat{\bm{\beta}}$ is the fitted coefficients of a given model, $df$ the number of non-zero elements in $\bm{\gamma}$, $\nu \in [0, 1]$ a constant, and $\sigma^{2}$ an unknown parameter analogue to the error variance in a standard regression model. We use  $\nu=\max\bracketlg[\big]{1-\frac{\log(n)}{2\log(pK)}, \frac{1}{2}}$ based on Theorem 1 and discussions in \cite{chen2008extended}. 
\cite{huang2010variable} and \cite{wang2015functional} substituted $n\sigma^{2}$ with residual sum of squares based on $\widehat{\bm{\beta}}$. Since a bridge estimator is not unbiased due to its shrinkage to zero, we instead propose to estimate $\sigma^{2}$ by the weighted residual sum of squares based on the generalized least-square estimator $\widehat{\bm{\beta}}_{GLS}$, which is unbiased although lacks sparsity. Letting $\widehat{\bm{\beta}}_{GLS} = (\bm{X}^{T}\bm{X})^{-1}\bm{X}^{T}\bm{Y}\bm{W}\bm{W}^{T}\bm{B}^{T}(\bm{B}\bm{W}\bm{W}^{T}\bm{B}^{T})^{-1}$ and 
$\widehat{{\sigma}}^{2}_{GLS} = \norm{(\bm{Y}-\bm{X}\widehat{\bm{\beta}}_{GLS})\bm{W}}_{2}^{2}/(nT),$ 
then the proposed adjusted EBIC is
\begin{equation}
        EBIC_{\nu} = T \frac{\norm{(\bm{Y}-\bm{X}\widehat{\bm{\beta}})\bm{W}}_2^2}{\norm{(\bm{Y}-\bm{X}\widehat{\bm{\beta}}_{GLS})\bm{W}}_{2}^{2}} + df \frac{\log{n}}{n} + \nu df \frac{\log{pK}}{n}, \label{eq:adjust.ebic}
\end{equation}
up to an additive constant. 
In the simulation, we observe that our proposed adjusted EBIC select parameters that are close to the oracle values; see Section \ref{s:simulation} for more details.

\section{Asymptotic Properties}\label{s:asymptotic-properties}

In this section, we study asymptotic properties of the proposed estimators. 
For a function $f(t): [0, 1] \rightarrow \mathbb{R}$, let $\|f\|_r$ be the $L_r$ norm and $\|f\|_{\infty} = \sup_{t \in [0, 1]} |f(t)|$. 
Denote by $\mathcal{C}^{r}[0,1]$ the H\"older space on $[0,1]$ with order $r$, a set of functions $f(\time)$ such that for some $L_f > 0$,
$|f^{(r_0)}(x) -  f^{(r_0)}(y)| \leq L_f \|x - y\|^{r - r_0}$ for all $x, y \in [0, 1]$, where $r_0$ is the largest integer strictly smaller than $r$. Let $S(f)=\{\time\in\mathcal{T}: f(\time)=0\}$ map $f(t)$ to its zero set. Denote by $\delta_{\min}(\bm{A})$ and $\delta_{\max}(\bm{A})$ the minimum and maximum eigenvalues for any given matrix $\bm{A}$. For two sequences $a_{n, T}$ and $b_{n, T}$, $a_{n, T} \lesssim b_{n, T}$ means $a_{n, T} \leq C b_{n, T}$ for some universal constant $C > 0$. We write $a_{n, T} \asymp b_{n, T}$ if $a_{n, T} \lesssim b_{n, T}$ and $a_{n, T} \gtrsim b_{n, T}$. Asymptotics in this section are interpreted when $n$ and $T$ go to infinity.  

We assume the following regularity conditions. 
\begin{assumption}
	\label{assumption:holder-cont} The underlying ${\beta}_{j}(\time) \in \mathcal{C}^{r}[0,1]$, for $j=1,2,\dots,p$ and $r\geq 2$.
\end{assumption}
\begin{assumption}
	\label{assumption:integral} The integral $\underset{t\in S({\beta}_{j})^c}{\int} \abs[s]{ \frac{1}{{\beta}_j(t)} }^{2(1 - \alpha)} d\time$ exists and is finite, for all $j$.
\end{assumption}
\begin{assumption}
$\bracketlg{{\epsilon}_{i}(\time)}_{i=1}^{n}$ and $\bracketlg{{\theta}_{i}(\time)}_{i=1}^{n}$ are independent across $i$ and sub-Gaussian.\label{assumption:iid-err}
\end{assumption}
\begin{assumption} 
The design matrix satisfies that for some constants $C_1, C_2 > 0$,  $C_1\leq\delta_{\min}(\bm{X}^T \bm{X}/n)$ and  $\delta_{\max}(\bm{X}^T \bm{X}/n) \leq C_2$, and the weight matrix is chosen such that $C_3 \leq \delta_{\min}(\bm{W} \bm{W}^T) \leq \delta_{\max}(\bm{W}\bm{W}^T) \leq C_4$ for some constants $C_3, C_4 > 0$, for all sufficiently large $n$ and $T$.\label{assumption:x-reg}
\end{assumption}
\begin{assumption}
	$\underset{\tidx}{\max}\bracketsm{\time_{\tidx+1}-\time_{\tidx}}=O(T^{-1})$. \label{assumption:equal-spaced}
\end{assumption}
\begin{assumption}
	$C_{5} K^{-1} \leq \time_{\tidx_{k+q}}-\time_{\tidx_{k}} \leq C_{6} K^{-1}$ for some constants $C_{5}, C_{6} > 0$, where $K<T$. \label{assumption:k}
\end{assumption}
{{Assumptions~\ref{assumption:holder-cont}, \ref{assumption:iid-err}, 
\ref{assumption:equal-spaced}, and \ref{assumption:k}, as well as the design matrix condition in Assumption \ref{assumption:x-reg}, are common in high-dimensional regression; see, for example, \cite{fan2000two, cai2011optimal, wang2015functional}. Assumptions \ref{assumption:equal-spaced} and \ref{assumption:k} are concerned with the spacing of time points and B-spline knots, respectively, and trivially hold when they are equally spaced.}}

Assumption \ref{assumption:integral} is in the same vein of Conditions (B') and (C') in~\cite{Fan2004a} to ensure that the group bridge penalty does not dominate the least square error on its support, and implies that ${\beta}_{j}(\time)$ deviates from zero fast enough so that  its support and zero set can be well distinguished. Unlike \cite{Fan2004a,huang2009group}, Assumption \ref{assumption:integral} disentangles the penalty and the regression function, leading to a simpler and more interpretable formulation to regularize regression coefficients. We achieve such simplicity by relying on a carefully modified B-spline approximation that will be detailed in Lemma~\ref{lem:appendix:approximation}. This assumption also suggests a lower bound of $\alpha$ for ${\beta}_{j}(\time)$ leaving zeros at polynomial speed. If ${\beta}_{j}(\time_{0})=0$ for some $\time_{0}$ and ${\beta}_{j}(\time)$ satisfies $\abs[s]{{\beta}_{j}(\time)} \gtrsim |\time-\time_{0}|^{b}$ as $t$ approaches $\time_{0}$ for some positive constant $b$, 
then choosing $\alpha> 1- 1/(2b)$ is required to comply with this assumption.

{{Assumption~\ref{assumption:x-reg} holds for the proposed data-driven $\bm{W}$ that satisfies $\bm{W}\bm{W}^T = \widehat{\bm{\Sigma}}^{-1} $ if the eigenvalues of $\widehat{\bm{\Sigma}}$ are bounded from above and below, a condition that is often satisfied following the rich literature of covariance estimation in functional data. For example, letting $C_{\bm{\theta}}$ be a constant such that $\delta_{\max}(\bm{\Sigma}_{\bm{\theta}}) \leq C_{\bm{\theta}}$, 
Theorem 2 of \cite{zhu2014spatially} proves the consistency of $\widehat{\bm{\Sigma}}^{-1}$, and particularly indicates $\lvert \delta_{\max}(\widehat{\bm{\Sigma}}_{\bm{\theta}}) - \delta_{\max}({\bm{\Sigma}}_{\bm{\theta}})\rvert \lesssim C_{\bm{\theta}}/2$ and $\lvert \widehat{\sigma}^{2} - {\sigma}^{2}\rvert \lesssim {\sigma}^{2}/2$, yielding $\lvert \delta_{\max}(\widehat{\bm{\Sigma}}) - \delta_{\max}({\bm{\Sigma}}) \rvert \lesssim (C_{\bm{\theta}}+{\sigma}^{2})/2$ and thus $\delta_{\max}(\widehat{\bm{\Sigma}}) \lesssim 3(C_{\bm{\theta}}+{\sigma}^{2})/2$. Moreover, the semi-positiveness of $\widehat{\bm{\Sigma}}_{\bm{\theta}}$ ensures that $\delta_{\min}(\widehat{\bm{\Sigma}})\geq \widehat{\sigma}^{2} \gtrsim {\sigma}^{2}/2$.}} 

It is well known that if ${\beta}_{j}(\time) \in \mathcal{C}^{r}[0,1]$ and there are no overlapping spline knots for $t_{m_{k}}$ for $q\leq k\leq K$, then there exists a B-spline approximation such that $L_{\infty}$ approximation error is upper bounded by $K^{-r}$ up to a constant; for example, see~\cite{schumaker2007spline}. However, these accurate B-spline approximations do not necessarily capture the sparsity of the true function. In Lemma~\ref{lem:appendix:approximation}, we propose a sparse modification of such B-splines so that the new approximation preserves the sparsity structure with the same approximation accuracy.  
\begin{lemma}
	\label{lem:appendix:approximation}Under Assumptions~\ref{assumption:holder-cont} and  \ref{assumption:k}, let $\widetilde{{\beta}}^{*}_{j}(\time)=\sum_{k=1}^{K}\widetilde{\bm{\gamma}}^{*}_{jk}{\phi}_{k}(\time)$ be the B-spline approximation in \cite{schumaker2007spline} such that $\big\|\widetilde{{\beta}}^{*}_{j}(\time) - {\beta}_{j}(\time)\big\|_{\infty} \leq C^{*}_{{\beta}_{j}}K^{-r}$ for a constant $C_{{\beta}_{j}}^{*}$. 
	Then there exists a sparse modification  $\widetilde{{\beta}}_{j}(\time):=\sum_{k=1}^{K}\widetilde{{\gamma}}_{jk}{\phi}_{k}(\time)$ and a constant $C_{{\beta}_{j}}$ such that $S({\beta}_{j}) \subseteq S(\widetilde{{\beta}}_{j})$, and $\big\|{\beta}_{j}(\time)-\widetilde{{\beta}}_{j}(\time)\big\|_{\infty} \leq C_{{\beta}_{j}} K^{-r}$. Also for any $\time\notin S(\widetilde{{\beta}}_{j})$, $\sum_{k:{\phi}_{k}(\time) > 0} \abs[s]{\widetilde{{\gamma}}_{jk}}\geq \frac{C_{{\beta}_{j}}^{*}}{qC_{{\beta}_{j}} + C_{{\beta}_{j}}^{*}}\abs[s]{{\beta}_{j}(\time)}$.
\end{lemma}
We call the modified B-spline coefficients $\widetilde{\bm{\gamma}} = \{\widetilde{{\gamma}}_{jk}\}_{p \times K}$ pseudo true values of $\bm{\gamma}$. The triangle inequality gives 
\begin{align}
        \big\|{\beta}_{j}(\time)-\widehat{{\beta}}_{j}(\time)\big\|_{\infty}  
    & \leq \big\|{\beta}_{j}(\time)-\widetilde{{\beta}}_{j}(\time)\big\|_{\infty} + \big\|\widetilde{{\beta}}_{j}(\time)-\widehat{{\beta}}_{j}(\time)\big\|_{\infty} 
    \\
    & \leq C_{{\beta}_{j}} K^{-r} + \big\|\widehat{\bm{\gamma}}_{j}-\widetilde{\bm{\gamma}}_{j}\big\|_{\infty}\sum_{k=1}^{K}\big\|{\phi}_{k}(\time)\big\|_{1}, 
\end{align}
where $\sum_{k=1}^{K}\norm{{\phi}_{k}(\time)}_{1}=1.$ Therefore, convergence rates of $\widehat{{\beta}}(\time)$ boils down to convergence rates of $\widehat{\bm{\gamma}}$ relative to the pseudo true $\widetilde{\bm{\gamma}}$ and the approximation error $C_{{\beta}_{j}} K^{-r}$ of using B-splines. The following Theorem~\ref{thm:5} and Theorem~\ref{thm:6} establish convergence rates of $\widehat{\bm{\gamma}}$ and $\widehat{{\beta}}_j$, respectively. 
To ease exposition, we relate $K$ and $T$ to $n$ by writing $K=n^{\frac{\kappa}{2r}}$ and $T=n^{\frac{\tau}{2r}}$ where $0<\kappa \leq\tau$. Here $n^{\frac{\kappa}{2r}}$ and $n^{\frac{\tau}{2r}}$ represent the asymptotic rates of $K$ and $T$, and any multiplicative constants do not change our results. To state the two theorems in their most general forms, we do not yet assume a specific order of either $\kappa$ or $\tau$. 

\begin{theorem}[Convergence rate of $\widehat{\bm{\gamma}}$]
\label{thm:5} Let $c_{\kappa}=\min(1, \kappa)$ and suppose Assumptions~\ref{assumption:holder-cont}--\ref{assumption:k} hold. If $\frac{\log(\lambda)}{\log(n)} \leq 1 - \frac{c_{\kappa}}{2} + \frac{\tau}{4r} - \frac{\kappa}{4r}$, then we have as $n\rightarrow\infty$
\begin{equation}
\norm{\widehat{\bm{\gamma}}-\widetilde{\bm{\gamma}}}_{\infty}\leq \norm{\widehat{\bm{\gamma}}-\widetilde{\bm{\gamma}}}_{2}=O_{p}\bracketsm[\big]{
n^{\frac{\kappa}{4r} - \frac{c_{\kappa}}{2}}
}. \label{eq:3-7}
\end{equation}
\end{theorem}

\begin{theorem}[Convergence rate of $\widehat{{\beta}}(\time)$]\label{thm:6}
Under the same assumptions of Theorem~\ref{thm:5}, as $n\rightarrow\infty$,
\begin{gather}
\big\|\widehat{{\beta}}_{j}(\time)-{\beta}_{j}(\time)\big\|_{\infty} =O_{p}
\bracketsm[\big]{n^{\frac{\kappa}{4r} - \frac{c_{\kappa}}{2}}},
\label{eq:6-2}\\
\big\|\widehat{{\beta}}_{j}(\time)-{\beta}_{j}(\time)\big\|_{2}  =O_{p}
\bracketsm[\big]{n^{ - \frac{c_{\kappa}}{2}}}. 
\label{eq:6-1}
\end{gather}
\end{theorem}

Theorem~\ref{thm:6} establishes convergence rates of $\widehat{{\beta}}(\time)$ under both $L_{2}$ and $L_{\infty}$ norms. We remark that there is a phase transition at $\tau = 1$ in \eqref{eq:6-2}. 
When $\tau < 1$, because $\kappa < \tau$, we have $c_{\kappa}=\kappa$, and the optimal $L_{\infty}$ rate is attained at the largest $\kappa$, i.e., $\kappa = \tau$, which gives the rate $O_p(n^{(1-2r)\frac{\tau}{4r}})$. In this case, the rate improves as $\tau$ increases.
When $\tau > 1$, the optimal $L_{\infty}$ rate is  $O_{p}(n^{(1 - 2r)\frac{1}{4r}})$, which is achieved at $\kappa = 1$, and increasing $\tau$ does not improve the rate. 
The same phase transition also applies to the $L_2$ rate in \eqref{eq:6-1}, which coincides with the observation made in~\cite{cai2011optimal}. In addition, our rate calculation implies that with $\kappa = \min(\tau, 1)$, the $L_2$ rate in \eqref{eq:6-1} becomes $O_p(n^{- \min(\tau, 1) / 2})$, i.e., $O_p(n^{-1/2} + T^{-r})$, which is minimax optimal~\citep{cai2011optimal}.

The rate under the $L_\infty$ norm in \eqref{eq:6-2} indicates that $\widehat{{\beta}}(\time)$ converges to ${\beta}(\time)$ at each $t$ for $\kappa < 2r$. 
This is particularly useful in detecting sparse regions as pointwise convergence suggests low false positive rates and low false negative rates in finding the support of ${\beta}_{j}(\time)$. In particular, 
we consider $\delta$-sparsity denoted by $S_{\delta}(h) = \{\time \in \mathcal{T}: |h(\time)| \leq \delta\}$ for $\delta > 0$. Then Equation~\eqref{eq:6-2} suggests that for arbitrary $\delta > 0$, as $n\rightarrow\infty$,
\begin{equation}
    P\{S(\widehat{{\beta}}_j) \subseteq S_{\delta}({\beta}_j)\} \rightarrow 1,\quad P\{S({{\beta}}_j) \subseteq S_{\delta}(\widehat{{\beta}}_j)\} \rightarrow 1
    .\label{eq:TPR}
\end{equation} 

{{In addition, the following theorem establishes one side of strict sparsity for the proposed method, that is, the support of $\widehat{{\beta}}(\time)$ is a subset of ${\beta}_{j}(\time)$ with probability approaching to $1$. This leads to low {false positive rates} under strict sparsity. It is an interesting future direction to study under what conditions the other side of strict sparsity also holds.}}

\begin{theorem}
\label{thm:9} Under the same conditions in Theorem~\ref{thm:5} 
and 
$\frac{\log(\lambda)}{\log(n)} > 1 + \frac{c_{\kappa}(\alpha-2)}{2} + \frac{\tau}{2r} - \frac{\kappa\alpha}{4r}$,
as $n\rightarrow\infty$, there holds 
$P\{S({\beta}_j) \subseteq S(\widehat{{\beta}}_j)\} \rightarrow 1.$

\end{theorem}

While we have focused on $\alpha <1$, a close inspection into the proofs of Theorems~\ref{thm:5} and \ref{thm:6} suggests that they also hold for $\alpha = 1$ (note that Assumption~\ref{assumption:integral} is not needed in this case). {{As discussed in Section~\ref{sec:group.penalty}, choosing $\alpha < 1$ induces exactly zero estimates of functional coefficients. Theorem \ref{thm:9} reassuringly shows that the recovered exactly zero regions tend to contain the zero regions of the true regression functions at least asymptotically. The proof of Theorem \ref{thm:9}
crucially relies on the choice of $\alpha < 1$; see the Appendix for details.}}  

\section{Simulations} \label{s:simulation}

We conduct simulations to compare finite sample performances of the proposed approach with competing methods in terms of function estimation and sparse region detection. We also assess the the proposed adjusted EBIC method for parameter tuning. 

\begin{figure}
	\centerline{
		\includegraphics[width=0.95\linewidth]{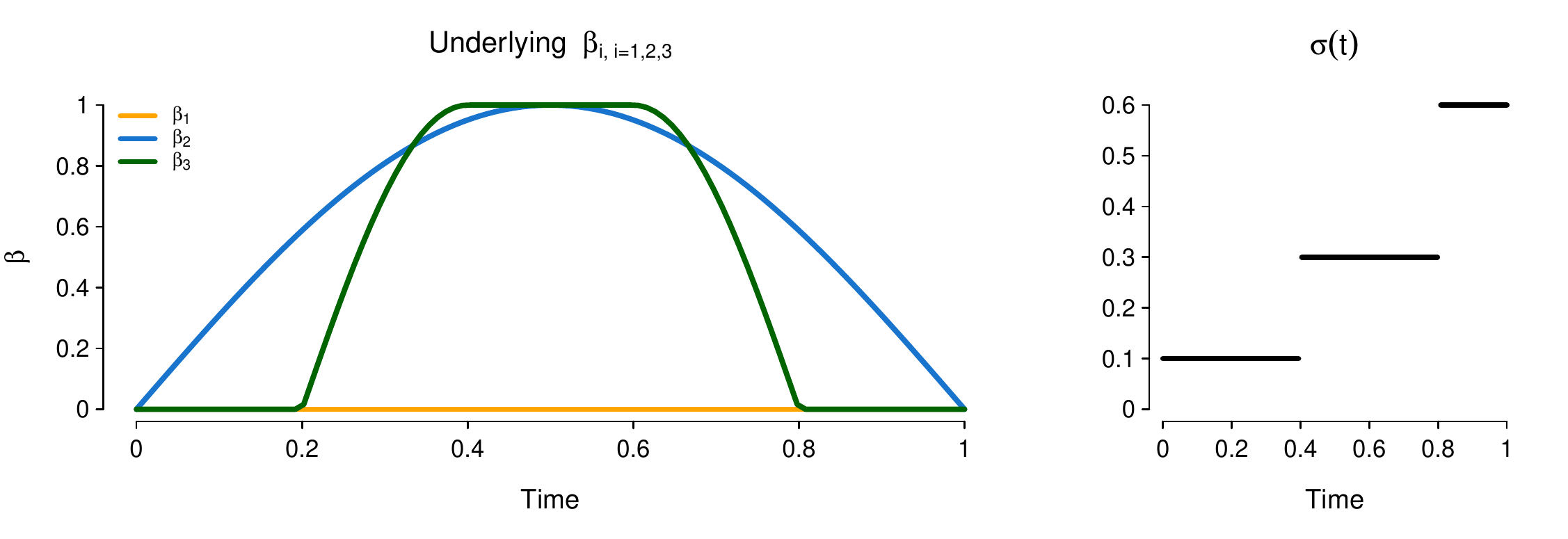}}
	\caption{Three coefficient functions (left) and step function $\sigma(\time)$ (right). ${\beta}_{1}(t)$ is globally sparse; ${\beta}_{2}(t)$ is dense; ${\beta}_{3}(t)$ is locally sparse. For ${\beta}_{3}(t)$, its phase I has the smallest errors and is half sparse and half dense. Phase II is dense with intermediate errors, and phase III is sparse with the largest errors.}
	\label{fig:truth}
\end{figure}

We generate data according to Model~\eqref{eq:FDA.model}. We consider three coefficients to represent different sparsity levels, displayed in Figure \ref{fig:truth}: ${\beta}_{1}(\time)=0$ for global sparsity, ${\beta}_{2}(\time)=\sin(\pi\time)$ for no sparsity (a dense coefficient), and ${\beta}_{3}(\time)$ for local sparsity that is defined as
\begin{equation}
{\beta}_{3}(t)=\sin\bracketsm{\frac{5\pi}{2}\time-\frac{\pi}{2}}\mathbf{1}_{[0.2, 0.4)}(t) + \mathbf{1}_{[0.4,0.6)}(t) + \sin\bracketsm{\frac{5\pi}{2}\time-\pi}\mathbf{1}_{[0.6,0.8)}(t).
\end{equation}
Among the three coefficient functions ${\beta}_3(t)$ is the most interesting one as it is locally sparse that pertains to the motivation of this paper; ${\beta}_1(t)$ and ${\beta}_2(t)$ are not of particular interest, but they allow us to study the performance of the proposed method when there are other coefficient functions with various sparsity levels. The design matrix $\bm{X}$ is generated from the standard normal distribution $N(0,1)$. To simulate different phases of experiments, the random effects ${\theta}_{i}(\time)$ are generated from a $AR(1)\times \sigma(\time)$ process, where $\sigma(\time)$ is a non-decreasing step function visualized in Figure~\ref{fig:truth} and $AR(1)$ is an order one autoregressive process with correlation $0.9$. 
The errors ${\epsilon}_i(\time_\tidx) \sim N(0,1)$ are independent across observations and time points. We use $T = 100$ as the time resolution to generate equally spaced $t_m$. We consider two sample sizes: $n = 100$ (small sample size) and $n = 1000$ (large sample size). We run 100 simulations for each sample size. 

In addition to the proposed weighted function group bridge approach, we include its two variants: homogeneous weight ($\bm{W}=\bm{I}$) and $\alpha=1$. Other competing methods include Group Lasso (gLasso) proposed by \cite{barber2017} and Group MCP (gMCP) by \cite{chen2016variable}. {{We also implement two-step function-on-scalar (2-Step FoS) regression \citep{fan2000two}, which first obtains regression coefficients at each time point then smooths these estimates. Although 2-Step FoS is not designed for functions with sparsity, we include it to compare estimation accuracy.}} For the proposed methods, we set $K = 30$ and choose $\lambda$ and $\alpha$, when applicable, by minimizing the adjusted EBIC in Equation~\eqref{eq:adjust.ebic} through a grid search using 100 $\lambda$'s chosen log-linearly from $0.1$ to $100$ and 18 $\alpha$'s linearly from $0.05$ to $0.95$. We derive joint confident bands for ${\beta}(t)$'s based on the variance estimation in Section \ref{s:variance}. In particular, we perturb sparse estimates $\widehat{{\gamma}}_{jk}$ with small random numbers to expand Equation~\eqref{eq:covariance} into all $\widehat{{\gamma}}_{jk}$'s, which gives the covariance of $\widehat{\bm{\gamma}}$ and subsequently a joint confident band for each ${\beta}(t)$.

\begin{figure}
	\centerline{
		\centering
		\includegraphics[width=0.95\linewidth, trim = 0 0 0 45, clip=TRUE ]{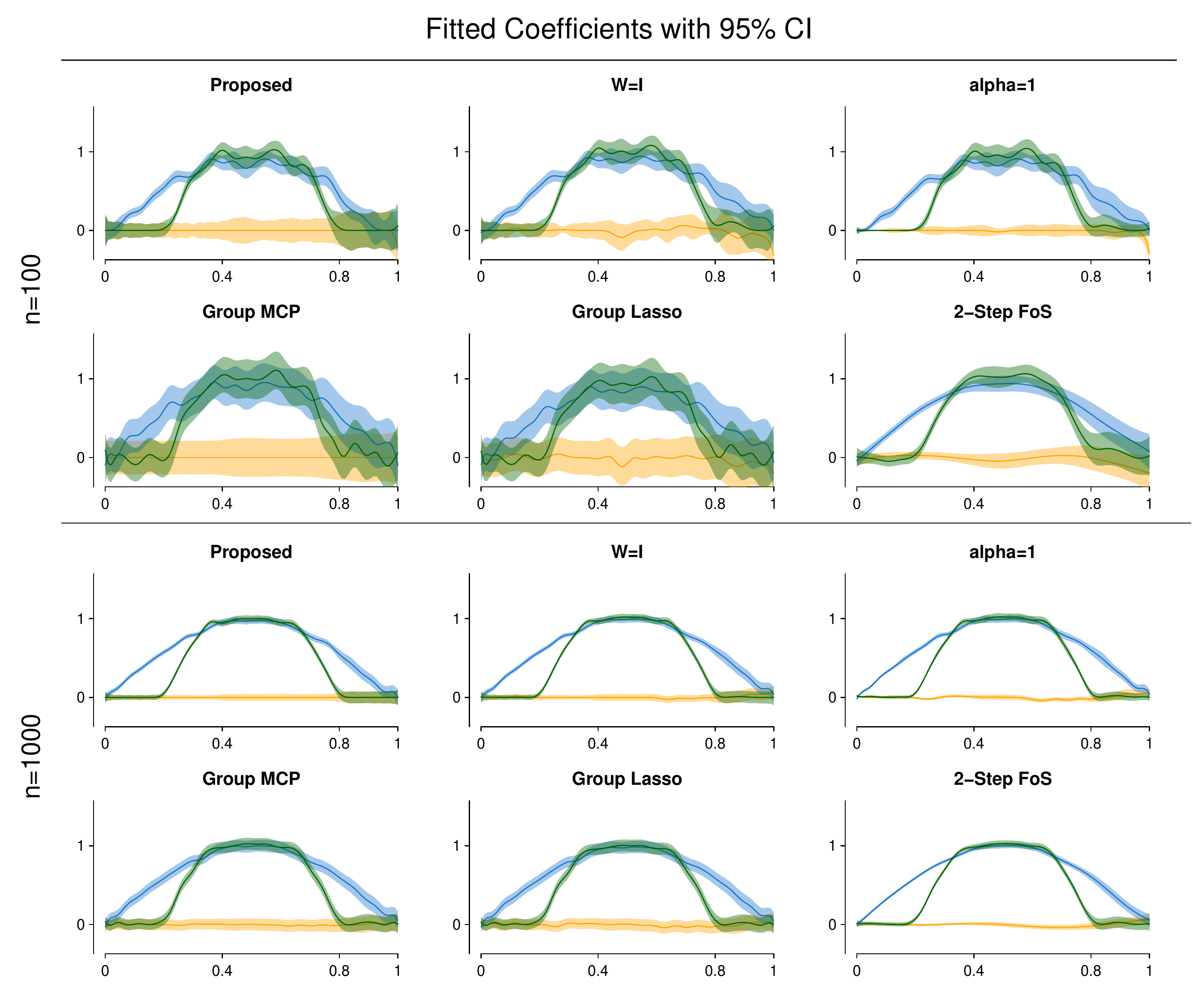}}
	\caption{
		\label{fig:all-coefs}
		Fitted coefficients with 95\% joint confidence intervals at $n = 100$ (top two rows) and $n = 1000$ (bottom two rows).
	}
\end{figure}

Figure \ref{fig:all-coefs} visualizes the estimates and their 95\% joint confidence intervals using one randomly selected replication. 
The proposed method appears to have tighter and more adaptive confidence intervals than gMCP and gLasso, partly due to its accounting for heterogeneous errors. Two-step FoS also enjoys tight joint confidence bands, but it cannot recover sparsity. Although all other methods contain sparsity constraints in their design, gLasso fails to detect the support of ${\beta}_{3}(\time)$. Functional group bridge methods and gMCP succeed in recovering the globally sparse signal ${\beta}_{1}(\time)$. The proposed method is the only method to recover the locally sparse coefficient ${\beta}_{3}(\time)$ at $\time\in (0.8,1]$, when the error variance is large, indicating its adaptivity to various sparse and noise levels. 

\begin{table}
 \centering
 \begin{minipage}{0.95\linewidth}
  \caption{
    Performance comparison of various methods. For each sample size $n \in \{100, 1000\}$, 
    RMSE calculates $\|{\beta}_{3}(\time) - \widehat{{\beta}}_{3}(\time) \|_{2}$, $L_{\infty}$ measures $\|{\beta}_{3}(\time) - \widehat{{\beta}}_{3}(\time) \|_{\infty}$, Coverage is the pointwise coverage of 95\% joint confidence intervals, and $F_1$ score assesses support recovery. 
    All results are averaged over 100 simulations. Standard errors are reported in parentheses. 
  }
  \label{tbl:rmse}
  \begin{tabular*}{0.9\linewidth}{@{}l@{\extracolsep{\fill}}r@{\extracolsep{\fill}}r@{\extracolsep{\fill}}r@{\extracolsep{\fill}}r@{\extracolsep{\fill}}r@{\extracolsep{\fill}}r@{\extracolsep{\fill}}r@{}}
  \hline
  \multirow{1}{*}{} $n$ &  Metrics & 
  Proposed & $\bm{W}=\bm{I}$
  & $\alpha=1$ & gMCP & gLasso & 2-Step FoS \\
 \hline
		\multirow{3}{*}{\rotatebox[origin=c]{90}{ $100$ }} 
		& \small{RMSE} \tiny{$(\times 0.01)$} &  8.6 (0.3) & \textbf{8.1} (0.3) & 8.8 (0.3) & 9.5 (0.2) & 10.2 (0.3) & 8.8 (0.2) \\ 
		& {$L_{\infty}$} \tiny{$(\times 0.01)$} &  \textbf{21.2} (0.9) & 23.0 (0.9) & 22.4 (0.7) & 26.2 (0.9) & 25.6 (0.7) & 23.1 (0.8) \\ 
		& {Coverage} &  94.8 (1.0) & 93.8 (0.9) & 87.3 (1.0) & 99.8 (0.1) & 99.5 (0.2) & 93.6 (0.6) \\ 
		& $F_{1}$ score & 0.85 (0.0) & 0.82 (0.0) & 0.79 (0.0) & 0.75 (0.0) & 0.75 (0.0) & 0.75 (0.0) \\
		\hline
		\multirow{3}{*}{\rotatebox[origin=c]{90}{ $1000$ }} 
		& \small{RMSE} \tiny{$(\times 0.01)$}  &  \textbf{2.4} (0.1) & 2.5 (0.1) & 3.2 (0.1) & 3.3 (0.1) & 3.6 (0.1) & 3.0 (0.1) \\ 
		& $L_{\infty}$ \tiny{$(\times 0.01)$} &  \textbf{7.6} (0.3) & 8.1 (0.3) & 8.5 (0.2) & 9.8 (0.3) & 9.7 (0.3) & 8.5 (0.3) \\ 
		& Coverage &  94.6 (0.4) & 96.4 (0.4) & 86.3 (0.8) & 99.6 (0.1) & 99.2 (0.2) & 94.6 (0.5) \\ 
		& $F_{1}$ score & 0.94 (0.0) & 0.90 (0.0) & 0.85 (0.0) & 0.75 (0.0) & 0.75 (0.0) & 0.75 (0.0) \\
		\hline
\end{tabular*}
\end{minipage}
\vspace*{6pt}
\end{table}

We next focus on ${\beta}_3(t)$ and compare each method in terms of both accuracy and support detection. For each method, we calculate the root mean squared error (RMSE) $\|\widehat{{\beta}}_3(\time) - {\beta}_3(\time)\|_{2}$ to measure the overall accuracy, and   $\|\widehat{{\beta}}_3(\time) - {\beta}_3(\time)\|_{\infty}$ to measure the maximum difference, reported in Table \ref{tbl:rmse}. We can see that our approach outperforms other methods in estimating the locally sparse coefficient ${\beta}_{3}(\time)$ for both sample sizes. 
Table \ref{tbl:rmse} also presents the coverage of the confident bands produced by each method. All methods except $\alpha=1$ attain the nominal coverage without significant deviation at $n = 1000$. Both gMCP and gLasso lead to the largest coverage at the expense of wider confidence bands; this can be clearly observed in Figure \ref{fig:all-coefs}. In contrast, the proposed method gives much tighter confidence bands while maintaining a coverage that is close to the nominal level. 

\begin{figure}
	\centering
	\includegraphics[width=\linewidth]{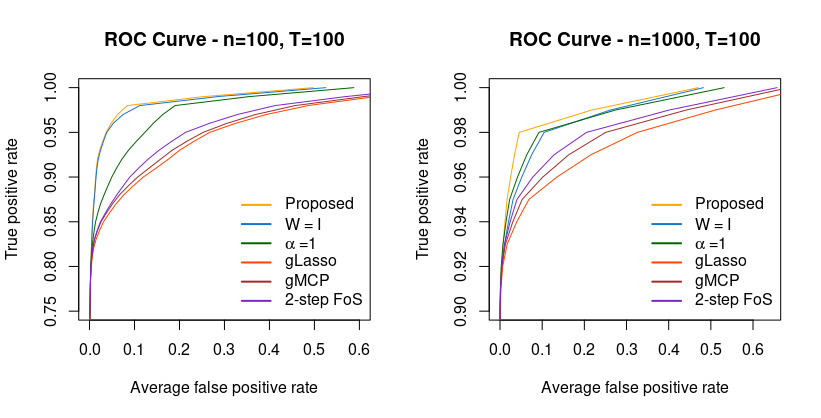}
	\caption{
		\label{fig:roc}
		ROC curve for each method with $n = 100$ (left) and $n = 1000$ (right), averaged over 100 simulations. 
	}
\end{figure}

{{For support detection, we calculate the false positive rate (FPR) or recall by $|S({\beta}_3) \cap S_{\delta}(\widehat{{\beta}}_3)^c|/|S({\beta}_3)|$, true positive rate (TPR) $|S({\beta}_3)^c \cap S_{\delta}(\widehat{{\beta}}_3)^c|/|S({\beta}_3)^c|$, and precision $|S({\beta}_3)^c \cap S_{\delta}(\widehat{{\beta}}_3)^c|/|S_{\delta}(\widehat{{\beta}}_3)^c|$ for some $\delta \geq 0$, where $|\cdot|$ counts the number of time points in an interval. Table \ref{tbl:rmse} reports the $F_{1}$ score $2/(\text{recall}^{-1} + \text{precision}^{-1})$ under strict sparsity ($\delta=0$), and 
Figure \ref{fig:roc} shows the receiver operating characteristic (ROC) curve by varying $\delta$, both averaged over 100 simulations.}} The proposed method gives the highest $F_1$ score for both sample sizes, corroborating sparsity recovery of functional group bridge. Additional results (in the Appendix) show that the proposed method leads to the lowest FPR, while the other three competing methods do not produce strict sparsity, thus yielding the same $F_1$ score. In Figure \ref{fig:roc}, group bridge-based methods, particularly the weighted version, tend to dominate other methods under $\delta$-sparsity. Overall, the substantial performance gain of the proposed method over $\alpha = 1$ and other methods may be partly due to the functional group bridge penalty for local sparsity and data-dependent weighting for heterogeneous volatility.  

\begin{figure}
	\centerline{
		\includegraphics[width=0.85\linewidth, trim = 75 25 45 25,]{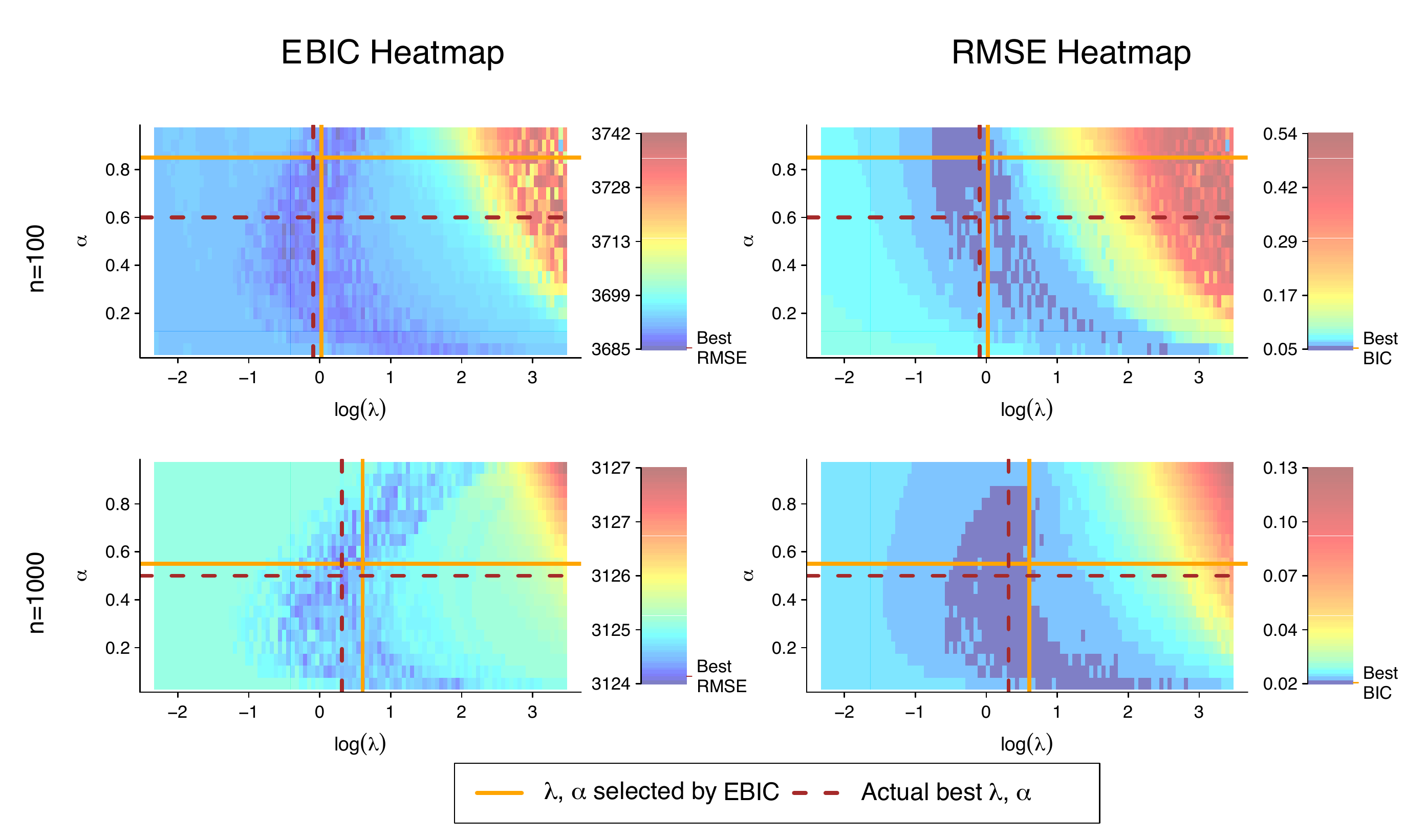}}
	\caption{
		\textbf{Left}: Adjusted EBIC heatmap with respect to combinations of $(\lambda, \alpha)$. 
		\textbf{Right}: Aggregated RMSE $\sum_{j = 1}^3 \norm{\widehat{{\beta}}_j(\time)-{\beta}_j(\time)}_{2}$. 
		Solid orange lines correspond to $(\lambda_{\mathrm{EBIC}}, \alpha_{\mathrm{EBIC}})$, the selected $(\lambda, \alpha)$ by the adjusted EBIC, and dash lines are $(\lambda_{\mathrm{RMSE}}, \alpha_{\mathrm{RMSE}})$, the optimal combination that minimizes RMSE. The small ticks on the legends: ``Best RMSE'' marks the EBIC value with $(\lambda_{\mathrm{RMSE}}, \alpha_{\mathrm{RMSE}})$; ``Best EBIC'' corresponds to the RMSE after using $(\lambda_{\mathrm{EBIC}}, \alpha_{\mathrm{EBIC}})$ to fit the model.
	}
	\label{fig:param-tuning}
\end{figure}

Figure~\ref{fig:param-tuning} assesses the proposed adjusted EBIC through a comparison with the aggregated RMSE $\sum_{j = 1}^3 \|\widehat{{\beta}}_j(\time)-{\beta}_j(\time)\|_{2}$. 
At $n = 1000$ the adjusted EBIC heatmap is highly consistent with the actual RMSE. The dark blue areas that indicate lowest EBIC and RMSE largely overlap, and the selected parameters are close to the actual best. For $n = 100$, although differing from the optimal ones, the selected $(\lambda, \alpha)$ give low RMSEs. 
Indeed, the optimal $(\lambda, \alpha)$ might not be unique as the RMSE heatmap shows a dark blue trajectory that achieves the best or close to the best accuracy.

\section{Application to iEEG Data} \label{s:application}

In this section, we apply the proposed method to a human intracranial electroencephalography (iEEG) dataset that is collected in Beauchamp's Lab to investigate multisensory integration and has been extensively described in \cite{ozker2018frontal} and \cite{karas2019cross}. 
In this experiment, participants either listened to recordings of words (auditory-only condition, A) or viewed videos and listened to recordings of words (audiovisual condition, AV). We are interested in the contrasting effect of A and AV on brain responses in  different brain areas. 

We focus on analyzing \textit{mouth-leading} words (e.g., ``last'' and ``drive''), meaning that mouth movements start before speech sounds. 
Such words were found to show a reduced brain response to audiovisual words compared to auditory-only words~\citep{karas2019cross}. {{The experiment contains 64 trials, each lasting for 3 seconds. There are 7 participating patients with a total of 58 Superior Temporal Gyrus (STG) electrodes. We fit the proposed model on each subject and electrode separately as the highly precise iEEG measurements localize activities of a small population of neurons nearest each electrode and lead to drastically different signals across electrodes. This separate analysis also allows us to study transferability of our findings. Information borrowing through jointly modelling all electrodes may yield further efficiency gain. This can be achieved by accounting for the spatial feature in a hierarchical model that links trials of various electrodes. The main challenges include the need to formulate the spatial dependence of signals and stochastic random effects across electrodes, and develop scalable algorithms for the increased parameter space.}} 

The original analog traces are measured at 2000 Hz. We apply notch filters to remove line noise and its harmonics (60, 120 and 180 Hz, etc.). Then a common average reference is used to remove common shifts introduced by patient activities. High-gamma oscillations usually stay above 70 Hz; hence, we apply wavelet transform to extract 70 - 150 Hz activities from the raw analogue traces. The transformed data is further down-sampled to 100 Hz for storage purposes. Each session is sliced into trials according to epoch information. All the trials are aligned to auditory onset, i.e., the time when audio stimuli started to emerge. Because there might be visual information before audio onset for mouth leading words, we collect three seconds of data for each trial, with one second prior and two seconds posterior to audio onset. Since brain activity levels often shift for each trial and frequency, we calibrate the signals of high-gamma activities against their own baselines (the average signals during the baseline period from $-1$ to $-0.3$ seconds). 
After the baseline period, we collapse the data by frequencies, resulting in a $301$ time-point functional data for each trial and electrode.

The functional response $\bm{Y}$ is an $64\times 301$ matrix for each electrode. The design matrix $\bm{X}$ is $64\times 2$, with the first column being constant one for the intercept and the second column indicating whether visual stimuli are present. The second regression coefficient ${\beta}_2(t)$ reflects the effect of audiovisual (AV) stimuli versus auditory only (A) stimuli, and  is of primary interest in this study. The time domain $\mathcal{T}$ is partitioned into four parts: $\mathcal{T}^{1}=[-1,-0.3)$ as the baseline period, $\mathcal{T}^{2}=[-0.3,0)$ containing video onset but without audio in, $\mathcal{T}^{3}=[0,1.5)$ when both auditory and visual stimuli are present, and $\mathcal{T}^{4}=[1.5,2]$ as clip offset. {{Since each trial is calibrated to the baseline, differential brain activities to experiment stimuli are expected to be zero during the baseline period, and non-zero when experiment stimuli present and exhibit effects.}} Consequently, the estimation of the locally sparse function ${\beta}_2(t)$ as well as detection of its support is of particular interest. 

\begin{figure}
	\centerline{
	    \includegraphics[width=0.95\linewidth]{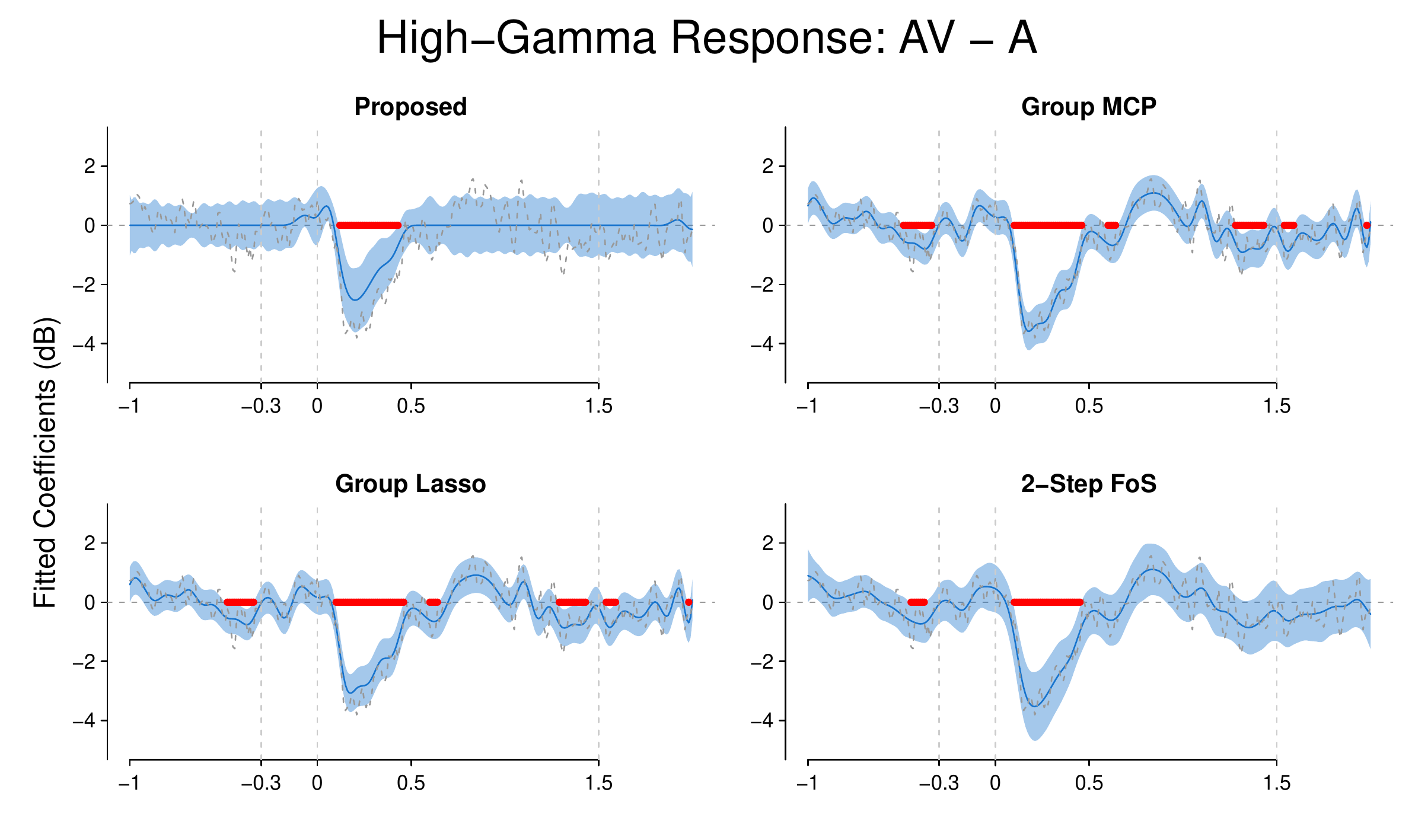}
	}
	\caption{Effect of AV versus A estimated by various methods. In each plot, the solid line (blue) is the estimated effect $\bm{\widehat{\beta}}_2(t)$, shaded area (light-blue) is the joint $95\%$ confidence band, and dashed line (grey) is the estimated effect from ordinary least squares at each time point. }
	\label{fig:e14-YAB}
\end{figure}

\begin{figure}
	\centering
	\includegraphics[width=0.85\linewidth]{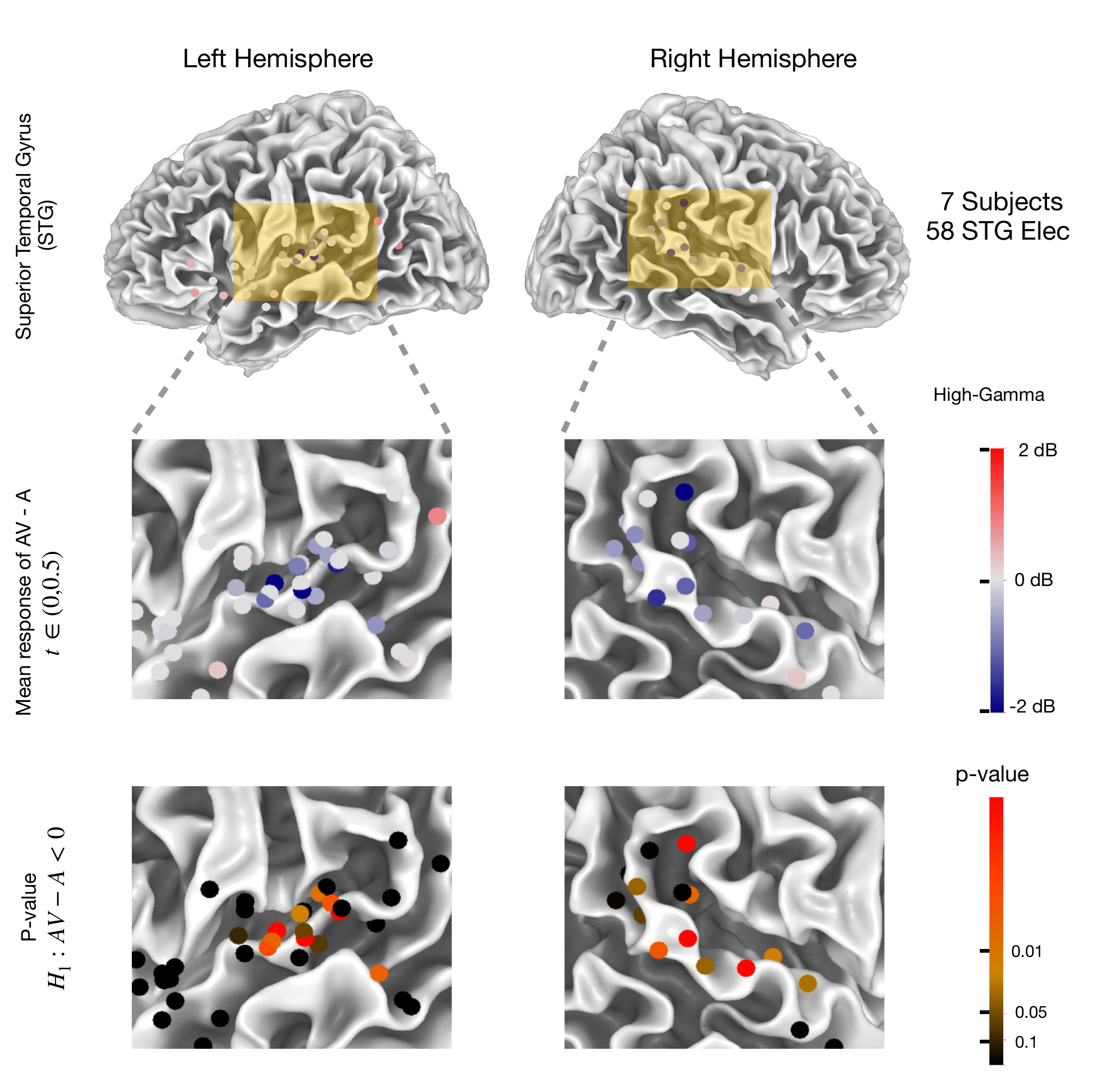}
	\caption{\textbf{First row}: visualization of all 58 electrodes mapped onto N27 template brain. \textbf{Second row}: mean response of AV-A from auditory onset to 500 ms after onset. Color coding: blue for suppression introduced by visual stimuli, gray for little to no differences, and red for that AV is greater than A. \textbf{Third row}: p-values for each electrodes with alternative hypothesis of AV less than A response. \label{fig:pstg}}
\end{figure}

We use $K=50$ when implementing the proposed method and select $\lambda$ and $\alpha$ by the adjusted EBIC. Figure \ref{fig:e14-YAB} plots the fitted coefficients for ${\beta}_2(t)$, AV versus A effect.  
In the proposed weighted functional group bridge approach, no significant signals are seen in the baseline window, while the other methods deviate from this expectation. 

According to \cite{karas2019cross}, we should expect a negative AV-A response after auditory onset as visual stimuli may suppress activities for words ``drive'' and ``last''. Because the proposed functional group bridge method is sparse on non-significant responses, it becomes easy not only to observe the suppression ($\text{AV}<\text{A}$), but to locate the starting time of that suppression as well as to automate the detection of duration of significant AV versus A effect. 
Figure \ref{fig:pstg} visualize all the 58 STG electrodes using the N27 template brain \citep{holmes1998enhancement}. 
The average response within 500 ms after auditory onset indicates that the posterior part of STG area is suppressed by additional visual stimuli when audio is present. To test the significance, we calculate the p-value to test the null hypothesis that there is no suppression ($\text{AV}\geq\text{A}$) for each electrode. The z-score is derived from $\min_{\time \in (0, 0.5)}\bracketlg[\big]{\widehat{{\beta}}(\time)/\widehat{s}(\time)}$, where $\widehat{s}(t)$ is the estimated standard derivation of $\widehat{{\beta}}(\time)$ as described in Section~\ref{s:simulation} based on \eqref{eq:covariance}. The corresponding p-values are displayed in the third row of Figure \ref{fig:pstg}. There are 15 electrodes in the posterior part of STG with p-values less than $0.05$. A closer inspection into the results show that each participant is associated with at least one significant electrode, indicating consistent visual surpressions in the posterior STG area for audiovisual stimuli. Our analysis for cross-modal suppression of auditory cortex complements the work by \cite{ozker2018frontal} and \cite{karas2019cross}, showing that multisensory interactions are a powerful modulator of activity throughout the speech perception network. Compared with the traditional methods used by \cite{ozker2018frontal, karas2019cross}, the proposed nonparametric method is more flexible with theroetical support. In addition, the proposed method provides a data-driven approach to find the time window where the brain response to each experimental condition differs, rather than relying on manually defined window as was done in the initial publications.

\section*{Acknowledgements}
This research was partly supported by the Grant DMS-2015569 from the National Science Foundation, and 1R24MH117529 from the BRAIN Initiative of the United States National Institutes of Health.

\bibliographystyle{apalike} \bibliography{main}

\newpage

\section{Appendix}

\subsection{Proofs}

In this appendix, we begin with a simple lemma related to B-splines that will be used in the subsequent proofs.

\begin{lemma}
	\label{lem:appendix:2}Suppose Assumptions \ref{assumption:equal-spaced} and \ref{assumption:k} hold and the B-spline order satisfies $q\geq 4$. Then for all $\bm{v}\in\mathbb{R}^{K}$ such that $\norm{\bm{v}}_{2}=1$, we have
	\begin{equation}
	    \norm[\bigg]{\sum_{k=1}^{K}v_{k}\phi_{k}(\time)}_{2}^{2} \asymp K^{-1},\ 
	    \frac{K}{T}\bm{v}^{T}\bm{B}\bm{B}^{T}\bm{v} \asymp 1.
	\end{equation}
\end{lemma}

\textit{Proof of Lemma \ref{lem:appendix:2}.} 
Let $\bm{D}$ be the diagonal matrix whose $k^{th}$ diagonal element is $(\time_{\tidx_{k+q}} - \time_{\tidx_{k}})/q$, where $q$ is the order of B-splines. According to Theorem 5.2 in \cite{de1976splines}, there exists a constant $C_q$ that only depends on $q$ such that
\begin{equation}
C_q^{-1}\norm{D^{1/2} \bm{v}}_{2} \leq \norm[\bigg]{\sum_{k=1}^{K}v_{k} \phi_{k}(\time) }_{2} \leq \norm{\bm{D}^{1/2} \bm{v}}_{2},\label{eq:appendix:1}
\end{equation}
which leads to
    $
        \norm[\big]{\sum_{k=1}^{K}v_{k}\phi_{k}(\time)}_{2}^{2}\asymp K^{-1}\norm{\bm{v}}_{2}^{2} = K^{-1}
    $
under Assumption~\ref{assumption:k}.

	Because B-splines are continuously differentiable of order $q - 2$, $\big\|\sum_{k=1}^{K}v_{k}\phi_{k}(\time)\big\|_{2}^{2}$ can be approximated by its Riemann sum
	\begin{equation}
	\norm[\bigg]{\sum_{k=1}^{K}v_{k}\phi_{k}(\time)}_{2}^{2}
	=\frac{1}{T}\norm{\bm{v}^{T}\bm{B}}_{2}^{2}+O\bracketlg{\underset{\tidx}{\max}\bracketsm{\time_{\tidx+1}-\time_{\tidx}}^{q-2}}
	=\frac{1}{T}\norm{\bm{v}^{T}\bm{B}}_{2}^{2}+O\bracketsm{T^{-q+2}}.
	\end{equation}
	With $K<T$ and $q\geq4$, we have
	\begin{equation}
	\frac{K}{T}\bm{v}^{T}\bm{B}\bm{B}^{T}\bm{v}  =K\norm[\bigg]{\sum_{k=1}^{K}v_{k}\phi_{k}(\time)}_{2}^{2}+O\bracketsm{KT^{-q+2}}
	\asymp\norm{\bm{v}}_{2}^{2}+O\bracketsm{K^{-q+3}}
	\asymp 1.
	\end{equation}

\textit{Proof of Lemma \ref{lem:appendix:approximation}.} 
We first consider a B-spline approximation by shrinking $\widetilde{\gamma}_{jk}^{*}$ such that $\abs[s]{\widetilde{\gamma}_{jk}^{*}} < \frac{C_{\beta_{j}}^{*}}{q}K^{-r}$ to zero. Define $\widetilde{\gamma}_{jk}^{**} = \widetilde{\gamma}_{jk}^{*}I(\abs[s]{\widetilde{\gamma}_{jk}^{*}} < \frac{C_{\beta_{j}}^{*}}{q}K^{-r})$, where $I(\cdot)$ is the indicator function, and $\widetilde{\beta}_j^{**}:=\sum_{k=1}^{K}\widetilde{\gamma}^{**}_{jk}\phi_{k}(\time)$ as the induced approximation. This new B-spline approximation satisfies that
\begin{equation}
    \norm{\widetilde{\beta}^{**}_{j}(\time) - \beta_{j}(\time)}_{\infty} 
    \leq
    \norm{\widetilde{\beta}^{*}_{j}(\time) - \beta_{j}(\time)}_{\infty} 
    + \sum_{k:\abs[s]{\widetilde{\gamma}_{jk}^{*}} < C_{\beta_{j}}^{*}K^{-r}/q}\abs{\widetilde{\gamma}_{jk}^{*}}\phi_{k}(\time)
    \leq 2C^{*}_{\beta_{j}}K^{-r}.
\end{equation}

Next, we partition B-spline knots as follows,
\begin{align}
	A_{j}^{1} & := A_{j}^1(\beta_{j}) = \bracketlg[\big]{k: \exists \time\in\bracketmd{\time_{\tidx_{k}}, \time_{\tidx_{k+1}}}\ \text{such that}\ \abs[s]{\beta_{j}(\time)} < C_{\beta_{j}}^{*}K^{-r}} \\
	A_{j}^{2} & := A_{j}^2(\beta_{j})= \bracketlg{k:k < k'\leq k+q, k'\in A_{j}^{1}
	} \\
	A_{j}^{3} & := A_{j}^3(\beta_{j})= \bracketlg{k:k\notin A_{j}^{1}\cup A_{j}^{2}}.
\end{align}
Let $S_j^{z} :=\bracketlg[\big]{\time:\time\in\bracketmd{\time_{\tidx_{k}},\time_{\tidx_{k+1}}},k\in A_{j}^{z}}$ for $z=1,2,3$. Then it is easy to show $S(\beta_{j})\subseteq S_j^{1}$. Also, for any $\time\in \bracketmd[s]{\time_{\tidx_{k}}, \time_{\tidx_{k+1}}} \subseteq S_{j}^{1}$, there exists at least one $\time' \in \bracketmd[s]{\time_{\tidx_{k}}, \time_{\tidx_{k+1}}}$ such that $\abs[s]{\beta_{j}(\time')} < C_{\beta_{j}}^{*}K^{-r}$. By the triangle inequality and $\max_{k}(\time_{\tidx_{k+1}} - \time_{\tidx_{k}}) \leq C_{6}K^{-1}$ as in Assumption \ref{assumption:k},
\begin{equation}
    \abs[s]{\widetilde{\beta}_{j}^{**}(\time)}\leq \abs[s]{\widetilde{\beta}_{j}^{**}(\time) - \beta_{j}(\time)} +  \abs[s]{\beta_{j}(\time')} + \abs[s]{\beta_{j}(\time) - \beta_{j}(\time')} 
    \leq \bracketsm[s]{3C_{\beta_{j}}^{*} + L_{\beta_{j}}C_{6}^{r}}K^{-r}. \label{eq:appendix:bound-pre-pseudo}
\end{equation}
We define refined sparse modifications $\widetilde{\gamma}$ and $\widetilde{\beta}_j(t)$ by
\begin{equation}
    \widetilde{\gamma}_{jk}=\widetilde{\gamma}_{jk}^{**}I\bracketlg[s]{k\in A_{j}^{3}},\quad
    \widetilde{\beta}_{j}(\time)=\sum_{k=1}^{K}\widetilde{\gamma}_{jk}\phi_{k}(\time).
\end{equation}
The modified $\widetilde{\beta}_j(t)$ preserves the sparsity of $\beta_{j}(\time)$. This is because $\bracketlg[\big]{k:\phi_{k}(\time)>0, \time\in S_{j}^{1}}\subseteq A_{j}^{1}\cup A_{j}^{2}$ according to construction of B-splines, yielding $\widetilde{\beta}_{j}(\time)=0$ for any $\time\in S_{j}^{1}$, i.e., $S(\beta_{j})\subseteq S(\widetilde{\beta}_{j})$. In addition, we can show $\norm{\beta_{j}(\time)-\widetilde{\beta}_{j}(\time)}_{\infty}\leq C_{\beta_{j}}K^{-r}$ for constant $C_{\beta_{j}}$. To see this,
we first note with $\abs[s]{\widetilde{\beta}_{j}^{**}(\time)-\widetilde{\beta}_{j}(\time)} \leq \max_{k}\abs[s]{\widetilde{\gamma}_{jk}^{**} - \widetilde{\gamma}_{jk}}\sum_{k=1}^{K}\phi_{k}(\time)$,
\begin{align}
    \abs[s]{\beta_{j}(\time)-\widetilde{\beta}_{j}(\time)}
    \leq \abs[s]{\beta_{j}(\time)-\widetilde{\beta}_{j}^{**}(\time)} +
    \abs[s]{\widetilde{\beta}_{j}^{**}(\time)-\widetilde{\beta}_{j}(\time)} 
    \leq 2C_{\beta_{j}}^{*}K^{-r} + \max_{k}\abs[s]{\widetilde{\gamma}_{jk}^{**} - \widetilde{\gamma}_{jk}},
\end{align}
For any $k$ such that $\widetilde{\gamma}_{jk}^{**} \neq \widetilde{\gamma}_{jk}$, $k\in A_{j}^{1}\cup A_{j}^{2}$ holds by construction. Thus, there always exists at least one $k'\in A_{j}^{1}$ such that $k\leq k' \leq k+q$. According to \eqref{eq:appendix:bound-pre-pseudo},  $\abs[s]{\widetilde{\beta}_{j}^{**}(\time')}\leq \bracketsm[s]{3C_{\beta_{j}}^{*} + L_{\beta_{j}}C_{6}^{r}}K^{-r}$ for any $\time'\in \bracketmd[s]{\time_{\tidx_{k'}}, \time_{\tidx_{k'+1}}}$. By the local support property of B-splines, there are at most $q$ B-splines that are non-zero at $\time'$, and particularly $\widetilde{\gamma}_{jk}^{**}\phi_{k}$ is one of them. In view of Theorem 5.2 in \cite{de1976splines}, $\abs[s]{\widetilde{\gamma}_{jk}^{**}}$ is upper bounded by $\bracketsm[s]{3C_{\beta_{j}}^{*} + L_{\beta_{j}}C_{6}^{r}}K^{-r}$ up to a constant. Therefore, there exists a constant $C_{\beta_{j}}$ such that $\norm{\beta_{j}(\time)-\widetilde{\beta}_{j}(\time)}_{\infty}\leq C_{\beta_{j}} K^{-r}$.

Finally, for all $\time\notin S(\widetilde{\beta}_{j})$, there exists at least one $k$ such that $\widetilde{\gamma}_{jk}\neq 0$ and $\phi_{k}(\time) > 0$. In this case, $\abs[s]{\widetilde{\gamma}_{jk}}=\abs[s]{\widetilde{\gamma}_{jk}^{**}}\geq C_{\beta_{j}}^{*}K^{-r}/q$. Hence,
\begin{equation}
    \frac{\sum_{k:\phi_{k}(\time)>0}\abs[s]{\widetilde{\gamma}_{jk}}}{\abs[s]{\beta_j(\time)}} 
    \geq \frac{\sum_{k:\phi_{k}(\time)>0}\abs[s]{\widetilde{\gamma}_{jk}}}{\abs[s]{\beta_j(\time) - \widetilde{\beta}_{j}(\time)} + \abs[s]{\widetilde{\beta}_{j}(\time)}} 
    \geq \frac{\sum_{k:\phi_{k}(\time)>0}\abs[s]{\widetilde{\gamma}_{jk}}}{C_{\beta_j}K^{-r} + \sum_{k:\phi_{k}(\time)>0}\abs[s]{\widetilde{\gamma}_{jk}}}
    \geq \frac{C_{\beta_{j}}^{*}}{qC_{\beta_{j}} + C_{\beta_{j}}^{*}}.
\end{equation}
This completes the proof.

\textit{Proof of Theorem \ref{thm:5}.} 
    Denote $a_{n,\kappa} := n^{\frac{\kappa}{4r} - \frac{c_{\kappa}}{2}}$, and let $\widetilde{\bm{\gamma}}$ be the sparse modified B-spline coefficients defined in Lemma \ref{lem:appendix:approximation}. 
	In order to prove $\norm{\widehat{\bm{\gamma}}-\widetilde{\bm{\gamma}}}_{2}=O_{p}\bracketsm[\big]{a_{n,\kappa}}$, 
	it is sufficient to show that, for arbitrary small $\epsilon>0$, there exists a universal constant $C$ such that 
	\begin{equation}
	P\bracketlg[\bigg]{\inf_{\norm{\bm{u}}_{2}=C}L\bracketsm[\Big]{\widetilde{\bm{\gamma}}+a_{n,\kappa}\bm{u}}>L\bracketsm{\widetilde{\bm{\gamma}}}}>1-\epsilon,\label{eq:proof.rate.prob}
	\end{equation}
	for all sufficiently large $n$. 
	
	Write $L\bracketsm[\big]{\widetilde{\bm{\gamma}}+a_{n,\kappa} \bm{u}}-L\bracketsm{\widetilde{\bm{\gamma}}}$ into
	\begin{equation}
	L\bracketsm[n]{\widetilde{\bm{\gamma}}+a_{n,\kappa}\bm{u}}-L\bracketsm{\widetilde{\bm{\gamma}}}=\bracketlg[\big]{f\bracketsm[n]{\widetilde{\bm{\gamma}}+a_{n,\kappa}\bm{u}}-f\bracketsm{\widetilde{\bm{\gamma}}}}+\lambda\bracketlg[\big]{g\bracketsm[n]{\widetilde{\bm{\gamma}}+a_{n,\kappa}\bm{u}}-g\bracketsm{\widetilde{\bm{\gamma}}}}.\label{eq:L.decompose} 
	\end{equation}
	We next bound the two terms on the right of~\eqref{eq:L.decompose} for all $\bm{u}$ such that $\|\bm{u}\|_2 = C$, where $C$ is to be determined later.  
	
	\textit{Lower bound of} $f\bracketsm[s]{\widetilde{\bm{\gamma}}+a_{n,\kappa} \bm{u}}-f\bracketsm{\widetilde{\bm{\gamma}}}$. For any $\bm{\gamma}$,
	\begin{equation}
	2f\bracketsm{\bm{\gamma}}=\norm{(\bm{Y}-\bm{X}\bm{\gamma} \bm{B})\bm{W}}_{2}^{2}=\sum_{i=1}^{n}\norm{\bracketsm{\bm{X}_{i}^{T}\bm{\gamma} \bm{B}-\bm{y}_{i}^{T}}\bm{W}}_{2}^{2}.
	\end{equation}
	Hence, 
	\begin{align}
	 &2f\bracketsm[s]{\bm{\gamma}+a_{n,\kappa}\bm{u}}-2f\bracketsm{\bm{\gamma}} 
	=\sum_{i=1}^{n}\norm[\big]{\bracketlg[\big]{\bm{X}_{i}^{T}\bracketsm[s]{\bm{\gamma}+a_{n,\kappa} \bm{u}}\bm{B}-\bm{y}_{i}^{T}}\bm{W}}_{2}^{2}-\norm[\big]{\bracketsm[s]{\bm{X}_{i}^{T}\bm{\gamma} \bm{B}-\bm{y}_{i}^{T}}\bm{W}}_{2}^{2}\\
	 =&\sum_{i=1}^{n}\bracketsm[\big]{\bm{X}_{i}^{T}a_{n,\kappa}\bm{u}\bm{B}+2\bm{X}_{i}^{T}\bm{\gamma} \bm{B}-2\bm{y}_{i}^{T}}\bm{W}\bm{W}^{T}\bracketsm[\big]{a_{n,\kappa} \bm{X}_{i}^{T}\bm{u}\bm{B}}^{T}\\
	 =&a_{n,\kappa}^{2}\sum_{i=1}^{n}\bm{X}_{i}^{T}\bm{u}\bm{B}\bm{W}\bm{W}^{T}\bm{B}^{T}\bm{u}^{T}\bm{X}_{i}+2a_{n,\kappa}\sum_{i=1}^{n}\bracketsm{\bm{X}_{i}^{T}\bm{\gamma} \bm{B}-\bm{y}_{i}^{T}}\bm{W}\bm{W}^{T}\bm{B}^{T}\bm{u}^{T}\bm{X}_{i}.
	 \label{eq:appendix:diff-fgamma}
	\end{align}
	Since $\bm{Y}=\bm{X}\bm{\gamma} \bm{B}+\bm{\theta}+\bm{X}\bm{R}+\bm{E}$, where $\bm{R}$ is B-spline approximation error, we have $\bm{X}_{i}^{T}\bm{\gamma} \bm{B}-\bm{y}_{i}^{T}=-\bm{X}_{i}^{T}\bm{R}-\bm{E}_{i}^{T}-\bm{\theta}_{i}$.
	Substituting this representation into  \eqref{eq:appendix:diff-fgamma} and letting $\bm{\gamma}=\widetilde{\bm{\gamma}}$ yields
	\begin{align}
	\label{eq:f.diff}
	2f\bracketsm[\big]{\widetilde{\bm{\gamma}}+a_{n,\kappa} \bm{u}}-2f\bracketsm{\widetilde{\bm{\gamma}}} = & a_{n,\kappa}^{2}\sum_{i=1}^{n}\bm{X}_{i}^{T}\bm{u}\bm{B}\bm{W}\bm{W}^{T}\bm{B}^{T}\bm{u}^{T}\bm{X}_{i}\\
	& -2a_{n,\kappa}\sum_{i=1}^{n}\bm{X}_{i}^{T}\bm{R}\bm{W}\bm{W}^{T}\bm{B}^{T}\bm{u}^{T}\bm{X}_{i} \\
	& -2a_{n,\kappa}\sum_{i=1}^{n}(\bm{E}_{i}+\bm{\theta}_{i})^{T}\bm{W}\bm{W}^{T}\bm{B}^{T}\bm{u}^{T}\bm{X}_{i} =: I_1 - 2I_2 - 2I_3.
	\end{align}
	The first term $I_1$ in  \eqref{eq:f.diff} can be bounded below by 
	\begin{align}
	    I_1 & \geq a_{n,\kappa}^{2}\frac{nT}{K}\delta_{\min}(\bm{W}\bm{W}^{T})\delta_{\min}
	    \left(\frac{K}{T}\bm{B}\bm{B}^{T}\right)\frac{1}{n}\sum_{i=1}^{n}\norm{\bm{X}_{i}^{T}\bm{u}}_{2}^{2}\\
	    & \geq a_{n,\kappa}^{2}\frac{nT}{K}\delta_{\min}(\bm{W}\bm{W}^{T})\delta_{\min}
	    \left(\frac{K}{T}\bm{B}\bm{B}^{T}\right) \sum_{i=1}^{n} \delta_{\min}\left(\frac{\bm{X}^{T}\bm{X}}{n}\right)C^{2}, \label{eq:bound.I1}
	\end{align}
	where the last step follows from 
	\begin{equation}
	   	\sum_{i=1}^{n}\norm{\bm{X}_{i}^{T}\bm{u}}_{2}^{2}  =\norm{\bm{X}^{T}\bm{u}}_{2}^{2}=\sum_{k=1}^{K}\norm{\bm{X}\bm{u}_{k}}_{2}^{2} \geq\delta_{\min}\bracketsm{\bm{X}^{T}\bm{X}}\sum_{k=1}^{K}\norm{\bm{u}_{k}}_{2}^{2}
	 = n\delta_{\min}\bracketsm{\bm{X}^{T}\bm{X}/n}C^{2} .
	\end{equation}
	We now substitute the eigenvalue conditions in Assumption \ref{assumption:x-reg} and Lemma \ref{lem:appendix:2} into~\eqref{eq:bound.I1} to obtain that
	\begin{equation}
	I_1 \geq C_0 a_{n,\kappa}^{2}\frac{nT}{K} C^{2}=C_{0}a_{n,\kappa}^{2}n^{1+\frac{\tau - \kappa}{2r}},\label{eq:thm5:3}
	\end{equation}
for some constant $C_0 > 0$ when $n$ is sufficiently large.

	We bound the second term $I_2$ in~\eqref{eq:f.diff} by
	\begin{align}
	I_2 = & a_{n,\kappa}\sum_{i=1}^{n}\bm{X}_{i}^{T}\bm{R}\bm{W}\bm{W}^{T}\bm{B}^{T}\bm{u}^{T}\bm{X}_{i} 
	\leq a_{n,\kappa}\sum_{i=1}^{n}\norm{\bm{X}_{i}^{T}}_{2}\norm{\bm{R}\bm{W}\bm{W}^{T}\bm{B}^{T}}_{2}\norm{\bm{u}}_{2}\norm{\bm{X}_{i}}_{2}\nonumber \\ 
	= & a_{n,\kappa}\sum_{i=1}^{n}\bm{X}_{i}^{T}\bm{X}_{i}C\norm{\bm{R}}_{2}\norm{\bm{W}\bm{W}^{T}}_{2}\norm{\bm{B}^{T}}_{2} \\
	\leq & a_{n,\kappa}n\delta_{\max}\bracketsm[\Big]{\frac{\bm{X}^{T}\bm{X}}{n}}\delta_{\max}(\bm{W}\bm{W}^{T})\norm{\bm{R}}_{2}\norm{\bm{B}}_{2} C \\ 
	= & a_{n,\kappa} O\bracketsm[\Big]{\frac{nT}{\sqrt{K}}K^{-r}} C
	= O\bracketsm[\Big]{a_{n,\kappa}^{2}n^{1+\frac{\tau - \kappa}{2r}}},
	  \label{eq:thm5:4} 
	\end{align}
	where \eqref{eq:thm5:4} is obtained by noticing $\norm{\bm{R}}_{2} = O\bracketsm[s]{K^{-r}T^{1/2}}$, $K^{-r+1/2}\leq a_{n,\kappa}$ by the definition of $a_{n,\kappa}$, and $\norm{\bm{B}}_{2}^{2}\asymp T/K$ by Lemma \ref{lem:appendix:2}. 
	
	For the third term $I_3$ in~\eqref{eq:f.diff}, we apply the Cauchy-Schwarz inequality to obtain 
	\begin{align}
	    & \sum_{i=1}^{n}(\bm{E}_{i}+\bm{\theta}_{i})^{T}\bm{W}\bm{W}^{T}\bm{B}^{T}\bm{u}^{T}\bm{X}_{i} 
	      = O_{p}\bracketmd[\Bigg]{ \delta_{\max}(\bm{B}\bm{W}\bm{W}^{T})\sqrt{\cov\bracketlg[\Big]{ \sum_{i=1}^{n}\sum_{t=1}^{T}(E_{it}^{2}+\theta_{it}^{2}) }}} C \\
	    & = O_{p}\bracketlg[\Big]{\sqrt{TK^{-1}}\sqrt{nT}\sqrt{\cov(E_{11}^{2}+\theta_{11}^{2})}}C = O_{p}\bracketlg[\Big]{n^{1+\frac{\tau - \kappa}{2r}}\sqrt{\cov(E_{11}^{2}+\theta_{11}^{2})}}C.
	\end{align}
According to the sub-Gaussianity condition in Assumption~\ref{assumption:iid-err}, we have $\cov(E_{11}^{2}+\theta_{11}^{2})=O_p(1)$. Therefore,
	\begin{equation}
	I_3 = a_{n,\kappa}\sum_{i=1}^{n}(\bm{E}_{i}+\bm{\theta}_{i})^{T}\bm{W}\bm{W}^{T}\bm{B}^{T}\bm{u}^{T}\bm{X}_{i}=
	O_{p} \bracketsm[\Big]{a_{n,\kappa}^{2}n^{1+\frac{\tau - \kappa}{2r}} }C.\label{eq:thm5:5}
	\end{equation}
	Combining (\ref{eq:thm5:3}), (\ref{eq:thm5:4}), and (\ref{eq:thm5:5}), we derive the following lower bound
	\begin{align}
	f\bracketsm[n]{\widetilde{\bm{\gamma}}+a_{n,\kappa}\bm{u}}-f\bracketsm{\widetilde{\bm{\gamma}}}
	& \geq a_{n,\kappa}^{2}n^{1+\frac{\tau - \kappa}{2r}} \bracketlg[\Big]{
	    \frac{C_0}{2}C^{2}- O(1)C - O_p(1)C
	}
	. \label{eq:thm5:6}
	\end{align}
	
	\textit{Lower bound of} $\lambda\bracketlg[\big]{g(\widetilde{\bm{\gamma}}+a_{n,\kappa}\bm{u})-g\widetilde{\bm{\gamma}}}$. We first bound $g_{j,m}\bracketsm[s]{\widetilde{\bm{\gamma}}+a_{n,\kappa}\bm{u}}-g_{j,m}\bracketsm{\widetilde{\bm{\gamma}}}$ by considering two cases as follows. If $\bgammajm{\widetilde{\bm{\gamma}}}_{1} > a_{n,\kappa} \bgammajm{\bm{u}}_{1}$, in view of the inequality
	$
	    \abs{y^\alpha - x^\alpha}= \abs{y^{\alpha-1} \cdot y - x^{\alpha-1} \cdot x}
	    \leq x^{\alpha-1} \abs{y - x}
	$
	for any $x,y>0$ and $\alpha\in (0,1]$, we obtain that
	\begin{align}
	    & \quad \abs{g_{j,m}\bracketsm[s]{\widetilde{\bm{\gamma}}+a_{n,\kappa}\bm{u}}-g_{j,m}\bracketsm{\widetilde{\bm{\gamma}}}} \\
	    &\leq \bgammajm{\widetilde{\bm{\gamma}}}_{1}^{\alpha-1} \abs{
	    \bgammajmC{\widetilde{\bm{\gamma}} + a_{n,\kappa} \bm{u}}_{1} - \bgammajm{\widetilde{\bm{\gamma}}}_{1}
	    }\\
	    &\leq \bgammajm{\widetilde{\bm{\gamma}}}_{1}^{\alpha-1} a_{n,\kappa} \bgammajm{\bm{u}}_{1}. 
	    \label{eq:lb-delta-g1}
	\end{align}
	If $0<\bgammajm{\widetilde{\bm{\gamma}}}_{1} \leq a_{n,\kappa}\bgammajm{\bm{u}}_{1}$, then
	\begin{align}
	    g_{j,m}\bracketsm[s]{\widetilde{\bm{\gamma}}+a_{n,\kappa}\bm{u}}-g_{j,m}\bracketsm{\widetilde{\bm{\gamma}}}
	    & \geq -g_{j,m}\bracketsm{\widetilde{\bm{\gamma}}} 
	    \geq -\bgammajm{\widetilde{\bm{\gamma}}}_{1}^{\alpha-1} a_{n,\kappa} \bgammajm{\bm{u}}_{1}.
	    \label{eq:lb-delta-g2}
	\end{align}
	Therefore, in both cases there holds $$g_{j,m}\bracketsm[s]{\widetilde{\bm{\gamma}}+a_{n,\kappa}\bm{u}}-g_{j,m} \geq -a_{n,\kappa}\bgammajm{\widetilde{\bm{\gamma}}}_{1}^{\alpha-1}  \bgammajm{\bm{u}}_{1}.$$ According to Lemma~\ref{lem:appendix:approximation}, either $\bgammajm{\widetilde{\bm{\gamma}}}_{1}=0$ or $\bgammajm{\widetilde{\bm{\gamma}}}_{1}\geq C_{1}\abs[s]{\beta_{j}(\time)}$ for some constant $C_{1}$. As a result, 
	\begin{align}
	& \lambda\sum_{j,\tidx}g_{j,m}\bracketsm[s]{\widetilde{\bm{\gamma}}+a_{n,\kappa}\bm{u}}-g_{j,m}\bracketsm{\widetilde{\bm{\gamma}}} \\
	\geq&  -\lambda a_{n,\kappa} \sum_{\underset{\abs[s]{\widetilde{\bm{\gamma}}_{j}}^{T}\mathbf{1}\{\bm{B}^{(m)}\}>0}{j,\tidx}}\bgammajm{\widetilde{\bm{\gamma}}}_{1}^{\alpha-1}  \bgammajm{\bm{u}}_{1} \\
	\geq& -\lambda a_{n,\kappa}
	\bracketmd[\bigg]{\sum_{\underset{\abs[s]{\widetilde{\bm{\gamma}}_{j}}^{T}\mathbf{1}\{\bm{B}^{(m)}\}>0}{j,\tidx}} \bgammajm{\widetilde{\bm{\gamma}}}_{1}^{2\alpha-2}}^{1/2} 
	\bracketmd[\bigg]{\sum_{j,\tidx}\bgammajm{\bm{u}}_{1}^{2}}^{1/2}\\
	\geq&    
	-\lambda C_{1}^{\alpha-1}\bracketlg[\bigg]{T\int_{\beta_{j}(\time)\neq 0}\beta_{j}(\time)^{2\alpha-2}d\time}^{1/2} q a_{n,\kappa}C
	\geq -C_{2}a_{n,\kappa}^{2}n^{1+\frac{\tau - \kappa}{2r}}C^{\alpha},
	\label{eq:lb-delta-g}
	\end{align}
	where $C_{2} = q \bracketlg{\int_{\beta_{j}(\time)\neq 0}\beta_{j}(\time)^{2\alpha-2}d\time}^{1/2}$
	and the last step in~\eqref{eq:lb-delta-g} follows the assumption that $\frac{\log(\lambda)}{\log(n)} \leq 1 - \frac{c_{\kappa}}{2} + \frac{\tau}{4r} - \frac{\kappa}{4r}$. 
	
	We now substitute \eqref{eq:thm5:6} and \eqref{eq:lb-delta-g} into~\eqref{eq:L.decompose} and obtain that 
	\begin{align}
	L\bracketsm[s]{\widetilde{\bm{\gamma}}+a_{n,\kappa} \bm{u}}-L\bracketsm{\widetilde{\bm{\gamma}}} 
	& \geq a_{n,\kappa}^{2}n^{1+\frac{\tau - \kappa}{2r}}\bracketlg[\Big]{\frac{C_0}{2}C^{2}-O_{p}(1)C - O(1)C^{\alpha} }.
	\label{eq:thm5:overall-lb}
	\end{align}
	Note that the constant $C_0$ does not depend on $n$. For arbitrary small $\epsilon > 0$, we can always choose a sufficiently large constant $C$ such that $\frac{C_0}{2} C^2 - O_{p}(1)C-O(1)C^{\alpha} > 0$ holds with probability at least $1 - \epsilon$ for sufficiently large $n$. Since $a_{n,\kappa}^{2}n^{1+\frac{\tau - \kappa}{2r}}\geq 1$ 
	for any $n>0$, the right hand side of \eqref{eq:thm5:overall-lb} is positive with probability at least $1 - \epsilon$, leading to~\eqref{eq:proof.rate.prob}. This completes the proof. 

\textit{Proof of Theorem \ref{thm:6}.} 
	We apply the triangle inequality
	$
	\norm{\widehat{\beta}_{j}-\beta_{j}}_{2}\leq\norm{\widehat{\beta}_{j}-\widetilde{\beta}_{j}}_{2}+\norm{\widetilde{\beta}_{j}-\beta_{j}}_{2}
	$ to decompose $\norm{\widehat{\beta}_{j}-\beta_{j}}_{2}$ into estimation error and approximation error. For $\time\in[0,1]$, there holds $\norm{\widetilde{\beta}_{j}-\beta_{j}}_{2}\leq\norm{\widetilde{\beta}_{j}-\beta_{j}}_{\infty}=O(K^{-r})$. Since $K^{-r}=n^{-\frac{\kappa}{2}}\leq n^{-\frac{c_{\kappa}}{2}}$, it suffices to calculate the rate for the dominating estimation error. 
	
	In view of Lemma \ref{lem:appendix:2}, we have $\norm{\widehat{\beta}_{j}-\widetilde{\beta}_{j}}_{2}^{2}\asymp n^{-\frac{\kappa}{2r}}\norm{\widehat{\bm{\gamma}}_{j}-\widetilde{\bm{\gamma}}_{j}}_{2}^{2},$
	which combined with   $\norm{\widehat{\bm{\gamma}}-\widetilde{\bm{\gamma}}}_{2}=O_p\bracketsm[\big]{n^{\frac{\kappa}{4r} - \frac{c_{\kappa}}{2}}}$ in Theorem \ref{thm:5} gives 
	\begin{equation}
	\norm{\widehat{\beta}_{j}-\widetilde{\beta}_{j}}_{2}\asymp n^{-\frac{\kappa}{4r}}\norm{\widehat{\bm{\gamma}}_{j}-\widetilde{\bm{\gamma}}_{j}}_{2}=O_p\bracketsm[\big]{n^{ - \frac{c_{\kappa}}{2}}}. \label{eq:approximation:l2-est-error}
	\end{equation}
	
	The $L_{\infty}$ norm rate is established in a similar manner by noting that   
	\begin{align}
	\norm{\widehat{\beta}_{j}-\widetilde{\beta}_{j}}_{\infty} & 
	\leq\norm{\widehat{\bm{\gamma}}-\widetilde{\bm{\gamma}}}_{\infty}
	\sum_{k=1}^{K}\norm{\phi_{k}(\time)}_{1}
	\leq\norm{\widehat{\bm{\gamma}}-\widetilde{\bm{\gamma}}}_{2}
	=O_p\bracketsm[\big]{n^{\frac{\kappa}{4r} - \frac{c_{\kappa}}{2}}}
	\end{align}
	as $\sum_{k=1}^{K}\phi_{k}(\time)=1.$ This completes the proof. 

\textit{Proof of Theorem \ref{thm:9}.} 
For each $\beta_{j}(\time)$, we apply the same partition to B-spline knots as in Lemma \ref{lem:appendix:approximation} and
define another sparse modification of the estimator $\widehat{\bm{\gamma}}'$ by $\widehat{\gamma}'_{jk} = \widehat{\gamma}_{jk}I\{k \in A_j^3\}$. Following a similar argument as in Lemma \ref{lem:appendix:approximation}, we can easily show that $\widehat{\beta}'(\time)=0$ when $\time \in S_{j}^{1}$.

	According to the KKT condition, there holds  $\frac{\partial}{\partial\bm{\gamma}_{j}}f(\widehat{\bm{\gamma}})+\sum_{j=1}^{p}\lambda\frac{\partial}{\partial\bm{\gamma}_{j}}s_{j,\tidx}(\widehat{\bm{\gamma}},\widehat{\zeta})=0$ for $j = 1, \ldots, p$.
	Expanding this derivative yields 
	\begin{equation}
	\bm{X}_{j}^{T}\bracketsm{\bm{Y} - \bm{X}\widehat{\bm{\gamma}}\bm{B}}\bm{W}\bm{W}^{T}\bm{B}^{T}= 
	\alpha\lambda\sum_{\tidx=1}^{T}\bgammajm{\widehat{\bm{\gamma}}}_{1}^{\alpha-1}\bracketmd[\Big]{\mathbf{1}\bracketlg[\big]{\bm{B}^{(m)}}\odot \text{sign}\bracketsm{\widehat{\bm{\gamma}}_{j}}}^{T}.
	\end{equation}
	Because either $\widehat{\gamma}_{jk}-\widehat{\gamma}'_{jk}=0$ (when $k\in A_{j}^{3}$) or $\widehat{\gamma}'_{jk}=0$ (when $k\notin A_{j}^{3}$), we have $\bracketsm[\big]{\widehat{\gamma}_{jk}-\widehat{\gamma}'_{jk}}\text{sign}\bracketsm{\widehat{\gamma}_{jk}}=\abs{\widehat{\gamma}_{jk}}I\bracketlg[\big]{k\notin A_{j}^{3}}$. In addition, $\big|\widehat{\gamma}'_{jk}\big|\leq \big|\widehat{\gamma}_{jk}\big|$ results in $\bgammajm{\widehat{\bm{\gamma}}'}_{1} \leq \bgammajm{\widehat{\bm{\gamma}}}_{1}$. Therefore,
	\begin{align}
	& \bm{X}_{j}^{T}\bracketsm{\bm{Y}-\bm{X}\widehat{\bm{\gamma}}\bm{B}}\bm{W}\bm{W}^{T}\bm{B}^{T}\bracketsm{\widehat{\bm{\gamma}}_{j}-\widehat{\bm{\gamma}}'_{j}}\\
	= & \alpha\lambda\sum_{\tidx=1}^{T}\bgammajm{\widehat{\bm{\gamma}}}_{1}^{\alpha-1}\times\sum_{k=1}^{K}\mathbf{1}\bracketsm[\big]{B_{k}^{(m)}}\abs{\widehat{\gamma}_{jk}}I\bracketlg[\big]{k\notin A_{j}^{3}}\\
	= & \alpha\lambda\sum_{\tidx=1}^{T}\bgammajm{\widehat{\bm{\gamma}}}_{1}^{\alpha-1} \bracketmd{\bgammajm{\widehat{\bm{\gamma}}}_{1}-\bgammajm{\widehat{\bm{\gamma}}'}_{1}}\\
	= & \alpha\lambda\sum_{\tidx=1}^{T}\bgammajm{\widehat{\bm{\gamma}}}_{1}^{\alpha} - \alpha\lambda\sum_{\tidx=1}^{T}\bgammajm{\widehat{\bm{\gamma}}}_{1}^{\alpha-1}\bgammajm{\widehat{\bm{\gamma}}'}_{1}\\
	\leq & \alpha\lambda\sum_{\tidx=1}^{T} \bracketlg[\big]{g_{j,\tidx}(\widehat{\bm{\gamma}}) - g_{j,\tidx}(\widehat{\bm{\gamma}}')} 
	= \alpha\lambda\sum_{\tidx:\time_{\tidx}\notin S_{j}^{1}} \bracketlg[\big]{g_{j,\tidx}(\widehat{\bm{\gamma}}) - g_{j,\tidx}(\widehat{\bm{\gamma}}')} + \alpha\lambda\sum_{\tidx:\time_{\tidx}\in S_{j}^{1}}g_{j,\tidx}(\widehat{\bm{\gamma}}) \\
	\leq & \lambda\sum_{\tidx:\time_{\tidx}\notin S_{j}^{1}} \bracketlg[\big]{g_{j,\tidx}(\widehat{\bm{\gamma}}) - g_{j,\tidx}(\widehat{\bm{\gamma}}')} + \alpha\lambda\sum_{\tidx:\time_{\tidx}\in S_{j}^{1}}g_{j,\tidx}(\widehat{\bm{\gamma}}).
	\label{eq:thm9:kkt}
	\end{align}
	As a result,
	\begin{align}
	    & \bm{X}_{j}^{T}\bracketsm{\bm{Y}-\bm{X}\widehat{\bm{\gamma}}\bm{B}}\bm{W}\bm{W}^{T}\bm{B}^{T}\bracketsm{\widehat{\bm{\gamma}}_{j}-\widehat{\bm{\gamma}}'_{j}} +(1-\alpha)\lambda \sum_{\tidx:\time_{\tidx}\in S_{j}^{1}}\bgammajm{\widehat{\bm{\gamma}}}_{1}^{\alpha} \\
	    \leq & \lambda\sum_{\tidx=1}^{T}\bracketmd{\bgammajm{\widehat{\bm{\gamma}}}_{1}^{\alpha}-\bgammajm{\widehat{\bm{\gamma}}'}_{1}^{\alpha}} = \lambda\sum_{\tidx=1}^{T} g_{j,m}(\widehat{\bm{\gamma}}) - \lambda \sum_{\tidx=1}^{T} g_{j,m}(\widehat{\bm{\gamma}}'),
	    \label{eq:thm9:kkt-final}
	\end{align}
	and consequently, 
		\begin{align}
	& \sum_{j=1}^{p}\bm{X}_{j}^{T}\bracketsm{\bm{Y}-\bm{X}\widehat{\bm{\gamma}}\bm{B}}\bm{W}\bm{W}^{T}\bm{B}^{T}\bracketsm{\widehat{\bm{\gamma}}_{j}-\widehat{\bm{\gamma}}'_{j}}+(1-\alpha)\lambda\sum_{j=1}^{p}\sum_{\tidx:\time_{\tidx}\in S_{j}^{1}}\bgammajm{\widehat{\bm{\gamma}}}_{1}^{\alpha}\\
	\leq & \lambda g(\widehat{\bm{\gamma}}) - \lambda g(\widehat{\bm{\gamma}}') \leq f(\widehat{\bm{\gamma}}') - f(\widehat{\bm{\gamma}}) 
= \frac{1}{2}\norm{\bracketsm{\bm{Y}-\bm{X}\widehat{\bm{\gamma}}'\bm{B}}\bm{W}}_{2}^{2}-\frac{1}{2}\norm{\bracketsm{\bm{Y}-\bm{X}\widehat{\bm{\gamma}}\bm{B}}\bm{W}}_{2}^{2}\\
	= & \frac{1}{2}\norm{\bm{X}\bracketsm{\widehat{\bm{\gamma}}-\widehat{\bm{\gamma}}'}\bm{B}\bm{W}}_{2}^{2}+\sum_{j=1}^{p}\bm{X}_{j}^{T}\bracketsm{\bm{Y}-\bm{X}\widehat{\bm{\gamma}}\bm{B}}\bm{W}\bm{W}^{T}\bm{B}^{T}\bracketsm{\widehat{\bm{\gamma}}_{j}-\widehat{\bm{\gamma}}'_{j}},
	\end{align}
	where the display in the second line uses the fact that $\widehat{\bm{\gamma}}$ minimizes the objective function $L(\bm{\gamma}) = f(\bm{\gamma}) + \lambda g(\bm{\gamma})$ in~\eqref{eq:gamma.hat}. It thus follows that 
	\begin{align}
	& (1-\alpha)\lambda\sum_{j=1}^{p}\sum_{\tidx:\time_{\tidx}\in S_{j}^{1}}\bgammajm{\widehat{\bm{\gamma}}}_{1}^{\alpha}  \leq\frac{1}{2}\norm{\bm{X}(\widehat{\bm{\gamma}}-\widehat{\bm{\gamma}}')\bm{B}\bm{W}}_{2}^{2}\\
	\leq & \frac{nT}{K}\delta_{\max}\bracketsm[\Big]{\frac{\bm{X}^{T}\bm{X}}{n}}\delta_{\max}\bracketsm[\Big]{\frac{K}{T}\bm{B}\bm{B}^{T}}\delta_{\max}(\bm{W}\bm{W}^{T})\norm{\widehat{\bm{\gamma}}-\widehat{\bm{\gamma}}'}_{2}^{2} = C_{0} n^{1+\frac{\tau-\kappa}{2r}} \norm{\widehat{\bm{\gamma}}-\widehat{\bm{\gamma}}'}_{2}^{2},
	\label{eq:thm9:combined}
	\end{align}
    by applying the eigen conditions in Assumption \ref{assumption:x-reg} and Lemma \ref{lem:appendix:2}. Because $0 < \alpha < 1$, the left hand side of the first line in \eqref{eq:thm9:combined} can be further lower bounded by
    \begin{align}
        \sum_{j=1}^{p}\sum_{\tidx:\time_{\tidx}\in S_{j}}\bgammajm{\widehat{\bm{\gamma}}}_{1}^{\alpha} 
        \geq \bracketmd[\Big]{\sum_{j=1}^{p}\sum_{\tidx:\time_{\tidx}\in S_{j}}\bgammajm{\widehat{\bm{\gamma}}}_{1}}^{\alpha}
        \geq \norm{\widehat{\bm{\gamma}} - \widehat{\bm{\gamma}}'}_{1}^{\alpha}.
        \label{eq:thm9:combined-lhs-lb}
    \end{align}
    Suppose $\norm{\widehat{\bm{\gamma}} - \widehat{\bm{\gamma}}'}_{2} > 0$, then combining \eqref{eq:thm9:combined} and \eqref{eq:thm9:combined-lhs-lb} gives 
    $
        (1-\alpha)\lambda \lesssim n^{1+\frac{\tau-\kappa}{2r}}\norm{\widehat{\bm{\gamma}} - \widehat{\bm{\gamma}}'}_{2}^{2-\alpha}.
    $ Hence
    \begin{equation}
        P(\norm{\widehat{\bm{\gamma}} - \widehat{\bm{\gamma}}'}_{2} > 0) 
        \leq P\bracketlg[\Big]{(1-\alpha)\lambda \leq C_{0} n^{1+\frac{\tau-\kappa}{2r}}\norm{\widehat{\bm{\gamma}} - \widehat{\bm{\gamma}}'}_{2}^{2-\alpha}}.
        \label{eq:thm9:p-bound}
    \end{equation}
	Noting that $\widehat{\gamma}'_{jk} = \widetilde{\gamma}_{jk} = 0$ for $k \notin A_{j}^{3}$ and  $\widehat{\gamma}'_{jk} = \widehat{\gamma}_{jk}$ for $k \in A_{j}^{3}$, we obtain
	\begin{align}
	\norm{\widehat{\bm{\gamma}}-\widehat{\bm{\gamma}}'}_{2}^{2}
	&	= \sum_{j=1}^{p}\bracketsm[\Big]{
		\sum_{k \notin A_{j}^{3}}\norm{\widehat{\gamma}_{jk}-\widehat{\gamma}'_{jk}}_{2}^{2}+
		\sum_{k \in A_{j}^{3}}\norm{\widehat{\gamma}_{jk}-\widehat{\gamma}'_{jk}}_{2}^{2}
	}\\
	&	= \sum_{j=1}^{p}\sum_{k \notin A_{j}^{3}}\norm{\widehat{\gamma}_{jk}-\widehat{\gamma}'_{jk}}_{2}^{2}
	= \sum_{j=1}^{p}\sum_{k \notin A_{j}^{3}}\norm{\widehat{\gamma}_{jk}-\widetilde{\gamma}_{jk}}_{2}^{2}
	\leq \norm{\widehat{\bm{\gamma}}-\widetilde{\bm{\gamma}}}_{2}^{2}=O_p(n^{\frac{\kappa}{4r} - \frac{c_{\kappa}}{2}})^2.
	\end{align}
    Substituting $\norm{\widehat{\bm{\gamma}}-\widehat{\bm{\gamma}}'}_{2} = O_p(n^{\frac{\kappa}{4r} - \frac{c_{\kappa}}{2}})$ into the right hand side of \eqref{eq:thm9:p-bound}, we have
    \begin{align}
        P(\norm{\widehat{\bm{\gamma}} - \widehat{\bm{\gamma}}'}_{2} > 0) 
        \leq P\bracketlg{ \lambda \leq O_p\bracketsm[s]{n^{1 + \frac{c_{\kappa}(\alpha-2)}{2} + \frac{\tau}{2r} - \frac{\kappa\alpha}{4r}}}}.
    \end{align}
    According to condition $\lambda \gtrsim n^{1 + \frac{c_{\kappa}(\alpha-2)}{2} + \frac{\tau}{2r} - \frac{\kappa\alpha}{4r}}$, one has $P(\norm{\widehat{\bm{\gamma}} - \widehat{\bm{\gamma}}'}_{2} > 0)\rightarrow 0$. As a result,
    \begin{equation}
        P\bracketlg[s]{S(\beta_{j}) \subseteq S(\widehat{\beta}_{j})}\rightarrow 1,
    \end{equation}
    as $n$ and $T$ go to infinity. This completes the proof.

\subsection{Additional Simulation Results}

In this section, we provide additional results that investigate the proposed method under strict sparsity, using the same simulation setting as in the main paper. 

Table~\ref{tbl:fpr2} reports the false positive rate (FPR) of each method under strict sparsity in the simulation study conducted in the main paper. We can see that the proposed method leads to the lowest FPR on all time segments that correspond to various noise levels, and the entire time domain $[0, 1]$, 
for both sample sizes. 
It is not surprising that 2-Step FoS does not recover sparse regions as it is not designed for sparse functions. Although gMCP and gLasso encourage sparsity, the estimated $\beta_{3}(\time)$ have no regions that are exactly zero. These observations align with Figure \ref{fig:all-coefs} in the main paper, suggesting excellent performance of the proposed method in support recovery for locally sparse functions. 
\begin{table}[ht]
	\centering
		\caption{
		\label{tbl:fpr2}
		False positive rates on estimating ${\beta}_{3}(\time)$ under strict sparsity. An estimate $\widehat{\beta}_{3}(\time)$ is considered false positive if $\widehat{\beta}_{3}(\time)\neq 0$ but ${\beta}_{3}(\time)= 0$.
	}
	\begin{tabular}{ l  l  l  llllll }
		\hline
		\multirow{1}{*}{} $n$ & Time & Proposed & $\bm{W}=\bm{I}$ & $\alpha=1$ & gMCP & gLasso & 2-Step FoS \\ 
		\hline
		\multirow{3}{*}{\rotatebox[origin=c]{90}{ $100$ }} 
		& 0 - 0.2 & \textbf{52.5} (2.4) & 69.1 (2.1) & 63.8 (2.2) & 99.8 (0.1) & 99.8 (0.1) & 100.0 (0.0) \\ 
		& 0.8 - 1 & \textbf{59.2} (3.3) & 63.1 (3.2) & 93.2 (1.3) & 100.0 (0.1) & 100.0 (0.1) & 100.0 (0.0) \\ 
		& Overall & \textbf{55.9} (2.1) & 66.1 (1.8) & 78.5 (1.2) & 99.9 (0.0) & 99.9 (0.1) & 100.0 (0.0) \\ 
		\hline
		\multirow{3}{*}{\rotatebox[origin=c]{90}{ $1000$ }} 
		& 0 - 0.2 & \textbf{15.2} (1.0) & 16.0 (1.2) & 36.1 (2.4) & 100.0 (0.0) & 100.0 (0.0) & 100.0 (0.0) \\
		& 0.8 - 1 & \textbf{45.8} (2.9) & 66.2 (2.8) & 84.2 (1.8) & 100.0 (0.0) & 100.0 (0.0) & 100.0 (0.0) \\ 
		& Overall & \textbf{30.4} (1.5) & 41.1 (1.5) & 60.1 (1.6) & 100.0 (0.0) & 100.0 (0.0) & 100.0 (0.0) \\ 
		\hline
	\end{tabular}
\end{table}

\subsection{Effect of $T$ and signal magnitude}
In this section, we carry out simulations to study the performance of the proposed method when the temporal resolution and signal-to-noise ratio vary. 

Case 1 is the same as Section~\ref{s:simulation} in the main paper with sample size $n = 100$, except that the number of time points $T$ is increased to 1000. This is motivated by high temporal resolution in iEEG study. Case 2 and Case 3 also adopt the settings in Section~\ref{s:simulation} in the main paper with sample size $n = 100$ but multiply the three regression coefficients by a factor $c_0 \in \{0.2, 5\}$, respectively. This investigates effects of signal magnitudes that control the signal-to-noise ratio on the performance of each method.

For the proposed methods, we focus on weighted functional group bridge with $\alpha < 1$ and omit its two variants with $\bm{W} = \bm{I}$ and $\alpha = 1$, since our experiments in the main paper do not suggest they achieve better performance. In Case 1, we set the number of B-splines to $K=40$ in light of the increased temporal resolution. Parameter tuning of the proposed method and three competing methods along with evaluation metrics follow Section~\ref{s:simulation} in the main paper unless stated otherwise. 

\begin{figure}[ht]
	\centerline{
		\centering
		\includegraphics[width=0.95\linewidth]{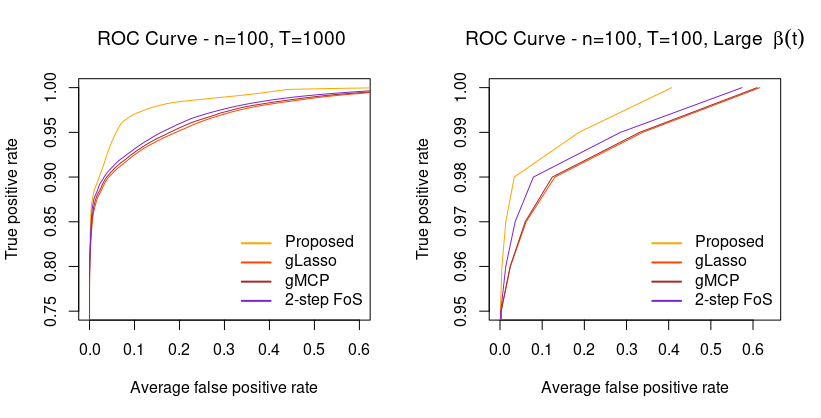}}
	\caption{ROC curve of each method in three cases, averaged over 100 simulations. \label{fig:sup-roc}
	}
\end{figure}

Table~\ref{tbl:sup-rmse} compares each method in terms of estimation accuracy and support detection under strict sparsity. We can see that the proposed method continues to give leading performance in both Case 1 and Case 2, achieving the smallest RMSE and $L_{\infty}$ distance while giving the largest $F_1$ score. The averaged ROC curves in Figure~\ref{fig:sup-roc} indicates superior performance of the proposed method in support detection under $\delta$-sparsity as the corresponding ROC curve tends to dominate the other three methods. These observations are consistent with Section~\ref{s:simulation} in the main paper. 

In Case 3, the maximal value of $\beta_3(t)$ as well as $\beta_2(t)$ decreases to 0.2. This poses daunting challenges to all approaches, including the proposed method. We observed that the $L_{\infty}$ metric $\|\widehat{\beta}_{3}(\time) - \beta_{3}(\time) \|_{\infty}$ averaged over 100 simulations for the four methods ranges from 1.8 to 2.2, which is way too large relative to the magnitude of $\beta_3(t)$. This is not surprising as the performance of each method is expected to deteriorate with extremely low signal-to-noise ratio.

\begin{table}
 \centering
 \begin{minipage}{0.95\linewidth}
  \caption{
    Performance comparison of various methods for estimating $\beta_3(t)$. For each case,
    RMSE calculates $\|\beta_{3}(\time) - \widehat{\beta}_{3}(\time) \|_{2}$, $L_{\infty}$ measures $\|\beta_{3}(\time) - \widehat{\beta}_{3}(\time) \|_{\infty}$, and $F_1$ score assess support recovery under strict sparsity. 
    All results are averaged over 100 simulations. Standard errors are reported in parentheses. 
  }
  \label{tbl:sup-rmse}
  \begin{tabular*}{0.9\linewidth}{@{}l@{\extracolsep{\fill}}r@{\extracolsep{\fill}}r@{\extracolsep{\fill}}r@{\extracolsep{\fill}}r@{\extracolsep{\fill}}r@{}}
  \hline
  \multirow{1}{*}{}  & Metrics & Proposed & gMCP & gLasso & 2-Step FoS \\
 \hline
    \multirow{4}{*}{\rotatebox[origin=c]{0}{ Case 1 }} 
	& \small{RMSE} \tiny{$(\times 0.01)$} &  \textbf{5.0} (0.1) & 7.0 (0.1) & 7.0 (0.1) & 6.3 (0.1) \\ 
	& {$L_{\infty}$} \tiny{$(\times 0.01)$} &  \textbf{18.1} (0.7) & 27.9 (0.9) & 27.6 (0.8) & 23.2 (0.8) \\
	& {$F_{1}$ score} & \textbf{0.94} (0.0) & 0.77 (0.0) & 0.77 (0.0) & 0.77 (0.0) \\
	\hline 
	\multirow{4}{*}{\rotatebox[origin=c]{0}{ Case 2 }} 
	& \small{RMSE} \tiny{$(\times 0.01)$} &  \textbf{8.1} (0.2) & 10.4 (0.2) & 11.6 (0.3) & 10.0 (0.2) \\
	& {$L_{\infty}$} \tiny{$(\times 0.01)$} &  \textbf{23.1} (0.6) & 30.1 (0.8) & 30.7 (0.8) & 28.7 (0.8) \\
	& {$F_{1}$ score} & \textbf{0.90} (0.0) & 0.76 (0.0) & 0.75 (0.0) & 0.76 (0.0) \\
	\hline 
\end{tabular*}
\end{minipage}
\vspace*{6pt}
\end{table}

\color{black}

\label{lastpage}

\end{document}